\documentclass[10pt,journal,compsoc]{IEEEtran}
\usepackage{cite}
\usepackage{amsmath}
\usepackage{amsfonts}
\usepackage{color}
\usepackage{comment}
\usepackage{subfigure}
\usepackage{graphicx}
\usepackage{multirow}
\usepackage{algorithmic}
\usepackage{bm}
\usepackage{url}
\usepackage{booktabs}  
\usepackage{bbding}

\newtheorem{Def}{Definition}

\begin{document}
\title{EKT: Exercise-aware Knowledge Tracing \\ for Student Performance Prediction}

\author{ Qi Liu, Zhenya Huang, Yu Yin, Enhong Chen, Hui Xiong,
	 Yu Su and Guoping Hu
	
	\IEEEcompsocitemizethanks{
		\IEEEcompsocthanksitem Q.~Liu, Z.~Huang, Y.~Yin and E.~Chen~(corresponding author) are with the Anhui Province Key Laboratory of Big Data Analysis and Application, School of Computer Science and Techonology, University of Science and Technology of China.
		\protect Email: \{huangzhy,yxonic\}@mail.ustc.edu.cn, \{qiliuql,cheneh\}@ustc.edu.cn
		
		\IEEEcompsocthanksitem H.~Xiong is with the Management Science and Information	Systems Department, Rutgers Business School, Rutgers, the State	University of New Jersey.
		\protect Email: hxiong@rutgers.edu.
		
		\IEEEcompsocthanksitem Y.~Su and G.~Hu are with iFLYTEK Research, IFLYTEK Co., Ltd, Hefei, Anhui 000000, China.
		\protect Email: \{yusu,gphu\}@iflytek.com.
		
	}
	
}
\markboth{IEEE TRANSACTIONS ON KNOWLEDGE AND DATA ENGINEERING}%
{Shell \MakeLowercase{\textit{et al.}}: Bare Advanced Demo of IEEEtran.cls for Journals}

\IEEEtitleabstractindextext{%
\begin{abstract}

For offering proactive services (e.g., personalized exercise recommendation) to the students in computer supported
intelligent education, one of the fundamental tasks is predicting student performance (e.g., scores) on future exercises, where it is necessary to track the change of each student's knowledge acquisition during her exercising activities. Unfortunately, to the best of our knowledge, existing approaches can only exploit the exercising records of students, and the problem of extracting rich information existed in the materials (e.g., knowledge concepts, exercise content) of exercises to achieve both more precise prediction of student performance and more interpretable analysis of knowledge acquisition remains underexplored. To this end, in this paper, we present a holistic study of student performance prediction. To directly achieve the primary goal of performance prediction, we first propose a general \emph{E}xercise-\emph{E}nhanced \emph{R}ecurrent \emph{N}eural \emph{N}etwork (EERNN) framework by exploring both student's exercising records and the text content of corresponding exercises. In EERNN, we simply summarize each student's state into an integrated vector and trace it with a recurrent neural network, where we design a bidirectional LSTM to learn the encoding of each exercise from its content. For making final predictions, we design two implementations on the basis of EERNN with different prediction strategies, i.e., \emph{EERNNM with Markov property} and \emph{EERNNA with Attention mechanism}. Then, to explicitly track student's knowledge acquisition on multiple knowledge concepts, we extend EERNN to an explainable \emph{E}xercise-aware \emph{K}nowledge \emph{T}racing (EKT) framework by incorporating the knowledge concept information, where the student's integrated state vector is now extended to a knowledge state matrix. In EKT, we further develop a memory network for quantifying how much each exercise can affect the mastery of students on multiple knowledge concepts during the exercising process. Finally, we conduct extensive experiments and evaluate both EERNN and EKT frameworks on a large-scale real-world data. The results in both general and cold-start scenarios clearly demonstrate the effectiveness of two frameworks in student performance prediction as well as the superior interpretability of EKT.

\end{abstract}

}

\maketitle


\IEEEpeerreviewmaketitle

\vspace{-0.6cm}
\section{Introduction} \label{sec:introduction}
\IEEEPARstart{I}{ntelligent} education systems, such as Massive Online Open Course, Knewton.com and KhanAcedemy.org, can help the personalized learning of students with computer-assisted technology by providing open access to millions of online courses or exercises. Due to their prevalence and convenience, these systems have attracted great attentions from both educators and general publics~\cite{anderson2014engaging,lan2014time-varying,zhao2018automatically}. 

\begin{figure*} [ht]
	\centering
	\includegraphics[scale=0.45]{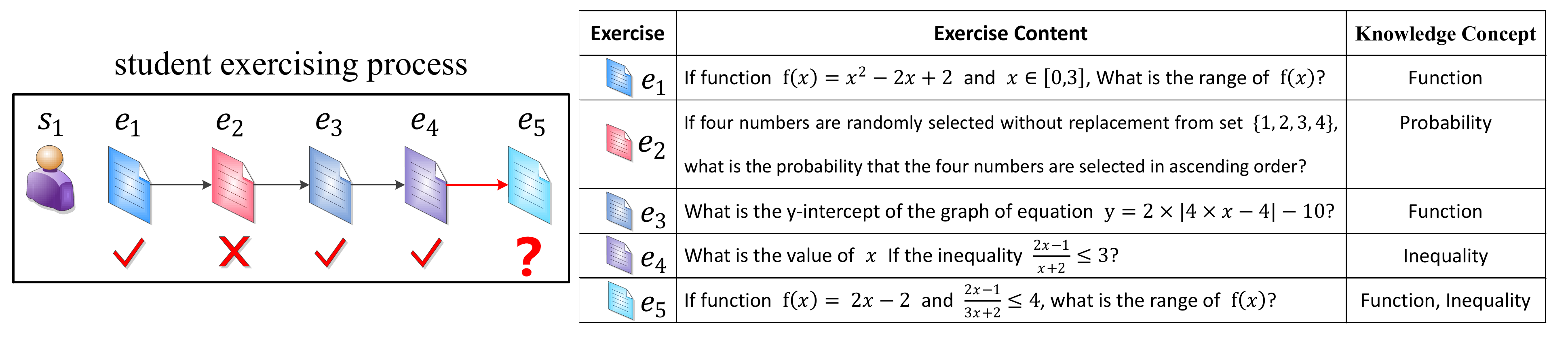}
	\vspace{-0.5cm}
	\caption{Example: Left box shows the exercising process of a student, where she has already done four exercises and is going to answer exercise $e_5$. Right table shows the corresponding materials of exercises that contain their contents and knowledge concepts.}
	\label{fig:IntroExample}\vspace{-0.6cm}
\end{figure*}

Specifically, students in these systems can choose exercises individually according to their needs and acquire necessary knowledge during exercising. Fig.~\ref{fig:IntroExample} shows a toy example of such exercising process of a typical student. Generally, when an exercise (e.g., $e_1$) is posted, the student reads its content (``If function...'') and applies the corresponding knowledge on ``Function'' concept to answer it. From the figure, student $s_1$ has done four exercises, where she only answers exercise $e_2$ wrong, which may demonstrate that she has well mastered knowledge concepts ``Function'' and ``Inequality'' except the ``Probability'' concept. We can see that a fundamental task in such education systems is to predict student performance (e.g., score), i.e., forecasting whether or not a student can answer an exercise (e.g., $e_5$) correctly in the future~\cite{baker2009state}. Meanwhile, it also requires us to track the change of students' knowledge acquisition in their exercising process~\cite{corbett1994knowledge,zhang2017dynamic}. In practice, the success of precise prediction could benefit both student users and system creators: (1) Students can realize their weak knowledge concepts in time and thus prepare targeted exercising~\cite{grossman2011transfer,wu2015cognitive}; (2) System creators can provide better proactive services to different students, such as learning remedy suggestion and personalized exercise recommendation~\cite{kuh2011piecing}.

In the literature, there are many efforts in predicting student performance from both educational psychology and data mining areas, such as cognitive diagnosis~\cite{dibello200631a}, knowledge tracing~\cite{corbett1994knowledge}, matrix factorization~\cite{thai2010recommender}, topic modeling~\cite{zhao2018automatically}, sparse factor analysis~\cite{lan2014time-varying} and deep learning~\cite{piech2015deep}. Specifically, existing works mainly focus on exploiting the exercising process of students, where each exercise is usually distinguished by the corresponding knowledge concepts in the modeling, e.g., exercise $e_1$ in Fig.~\ref{fig:IntroExample} is represented as the concept ``Function''. In other words, existing works model students' knowledge states for the prediction only based on their performance records on each knowledge, where two exercises (e.g., $e_1$ and $e_3$) labeled with the same knowledge concept are simply identified as the same (actually, exercise $e_1$ and $e_3$ are quite different according to their contents, and $e_3$ is more difficult than $e_1$). Therefore, these approaches cannot distinguish the knowledge acquisition of two students if one solves $e_1$ but the other solves $e_3$ since these knowledge-specific representations underutilize the rich information of exercise materials (e.g., text contents), causing severe information loss~\cite{dibello200631a}. To this end, we argue that it is beneficial to combine both student's exercising records and the exercise materials for more precisely predicting student performance.

Unfortunately, there are many technical and domain challenges along this line. First, there are diverse expressions
of exercises, which requires a unified way to automatically understand and represent the characteristics of exercises from a semantic perspective. Second, students' performance in the future is deeply relied on their long-term historical exercising, especially on their important knowledge states. How to track the historically focused information of students is very challenging. Third, the task of student performance prediction usually suffers from the ``cold start" problem~\cite{liu2014a,wilson2016estimating}. That is, we have to make predictions for new students and new exercises. In this scenario, limited information could be exploited, and thus, leading to the poor prediction results. Last but not least, students usually care about not only what they need to learn but also wonder why they need it, i.e., it is necessary to remind them whether or not they are good at a certain knowledge concept and how much they have already learned about it. However, it is a nontrivial problem to either quantify the impacts of solving each specific exercise (e.g., $e_1$) on improving the student's knowledge acquisition (e.g., ``Function'') or interpretably track the change of student's knowledge states during the exercising process.

To directly achieve the primary goal of predicting student performance with addressing the first three challenges, in our preliminary work~\cite{Su2018exercise}, we proposed an \emph{E}xercise-\emph{E}nhanced \emph{R}ecurrent \emph{N}eural \emph{N}etwork (EERNN) framework by mainly exploring both student's exercising records and the corresponding exercise contents. Specifically, for the exercising process modeling, we first designed a bidirectional LSTM to represent the semantics of each exercise by exploiting its content. The learned encodings could capture the individual characteristics of each exercise without any expertise. Then, we proposed another LSTM architecture to trace student states in the sequential exercising process with the combination of exercise representations. For making final predictions, we designed two strategies on the basis of EERNN framework. The first one was a straightforward yet effective strategy, i.e., \emph{EERNNM with Markov property}, in which the students' next performance only depended on current states. Comparatively, the second was a more sophisticated one, \emph{EERNNA with Attention mechanism}, which tracked the focused student states based on similar exercises in the history. In this way, EERNN could naturally predict student's future performance given her exercising records.

In EERNN model, we summarized and tracked each student's knowledge states on all concepts in one integrated hidden vector. Thus, it could not explicitly explain how much she had mastered a certain knowledge concept (e.g., ``Function''), which meant that the interpretability of EERNN was not satisfying enough. Therefore, in this paper, we extend EERNN and propose an explainable \emph{E}xercise-aware \emph{K}nolwedge \emph{T}racing (EKT) framework to track student states on multiple explicit concepts simultaneously. Specifically, we extend the integrated state vector of each student to a knowledge state matrix that updates over time, where each vector represents her mastery level of a certain concept. At each exercising step of a certain student, we develop a memory network to quantify the different impacts on each knowledge state when she solves a specific exercise. We also implement two EKT based prediction models following the proposed strategies in EERNN, i.e., \emph{EKTM with Markov property} and \emph{EKTA with Attention mechanism}. Finally, we conduct extensive experiments and evaluate both EERNN and EKT frameworks on a large-scale real-world dataset. The experimental results in both general and cold-start scenarios clearly demonstrate the effectiveness of two proposed frameworks in student performance prediction as well as the superior interpretability of EKT framework.

\section{Related Work}
The related work can be classified into following categories from both educational psychology (i.e., cognitive diagnosis and knowledge tracing) and data mining (i.e., matrix factorization and deep learning methods) areas.

\textbf{Cognitive Diagnosis.}
In the domain of educational psychology, cognitive diagnosis is a kind of techniques that aims to predict student performance by discovering student states from the exercising records~\cite{dibello200631a}. Generally, traditional cognitive diagnostic models (CDM) could be grouped into two categories: continuous models and discrete ones. Among them, item response theory (IRT), as a typical continuous model, characterized each student by a variable, i.e., a latent trait that describes the integrated knowledge state, from a logistic-like function~\cite{embretson2013item}. Comparatively, discrete models, such as \emph{Deterministic Inputs, Noisy-And gate model} (DINA), represented each student as a binary vector which denoted whether she mastered or not the knowledge concepts required by exercises with a given Q-matrix (exercise-knowledge concept matrix) prior~\cite{de2009dina}. To improve prediction effectiveness, many variations of CDMs were proposed by combining learning information~\cite{cen2006learning,pavlik2009performance,wu2015cognitive}. For example, learning factors analysis (LFA)~\cite{cen2006learning} and performance factors analysis (PFA)~\cite{pavlik2009performance} incorporated the time factor into the modeling. Liu et al.~\cite{liu2018fuzzy} proposed FuzzyCDM that considered both subjective and objective exercise types to balance precision and interpretability of the diagnosis results.

\textbf{Knowledge Tracing.}
Knowledge tracing is an essential task for tracing the knowledge states of each student separately, so that we can predict her performance on future exercising activities, where the basic idea is similar to the typical sequential behavior mining~\cite{liu2017multi-behavioral,shang2017searching}. In this task, Bayesian knowledge tracing (BKT)~\cite{corbett1994knowledge} was one of the most popular models. It was a knowledge-specific model which assumed each student's knowledge states as a set of binary variables, where each variable represented she had ``mastered'' or ``non-mastered'' on a specific concept. Generally, BKT utilized a Hidden Markov Model~\cite{rabiner1986introduction} to update knowledge states of each student separately followed by her performance on exercises. On the basis of BKT, many extensions were proposed by considering other factors, e.g., exercise difficulty~\cite{pardos2011kt}, multiple knowledge concepts~\cite{xu2010using} and student individuals~\cite{yudelson2013individualized}. One step further, to improve the prediction performance, other researchers also suggested to incorporate some cognitive factors into traditional BKT model~\cite{khajahincorporating,khajah2014integrating}.

\textbf{Matrix Factorization.}
Recently, researchers have attempted to leverage matrix factorizations from data mining field for student performance prediction~\cite{toscher2010collaborative,thai2010recommender}. Usually, the goal of this kind of research is to predict the unknown scores of students as accurate as possible given a student-exercise performance matrix with some known scores. For example, Thai et al.~\cite{thai2010recommender} leveraged matrix factorization models to project each student into a latent vector that depicted students' implicit knowledge states, and further proposed a multi-relational adaption model for the prediction in online learning systems. To capture the changes of student's exercising process, some additional factors are incorporated. For example, Thai et al.~\cite{thai2011factorization} proposed a tensor factorization approach by adding additional time factors. Chen et al.~\cite{chen2014tracking} noticed the effects of both Learning theory and Ebbinghaus forgetting curve theory and incorporated them into a unified probabilistic framework. Teng et al.~\cite{teng2018interactive} further investigated the effects of two concept graphs.

\textbf{Deep Learning Methods.}
Learning is a very complex process, where the mastery level of students on different knowledge concepts is not updated separately but related to each other. Along this line, inspired by the remarkable performance of deep learning techniques in many applications, such as speech recognition~\cite{graves2013speech}, image learning~\cite{krizhevsky2012imagenet,cui2017general}, natural language processing~\cite{mikolov2013distributed}, network embedding~\cite{cui2018survey,zhu2018deep}, and also educational applications like question difficulty prediction~\cite{huang2017question}, some researchers attempted to use deep models for student performance prediction～\cite{piech2015deep,zhang2017dynamic}. Among these work, deep knowledge tracing (DKT) was the first attempt, to the best of our knowledge, to utilize recurrent neural networks (e.g., RNN and LSTM) to model student's exercising process for predicting her performance~\cite{piech2015deep}. Moreover, by bridging the relationship between exercises and knowledge concepts, Zhang et al.~\cite{zhang2017dynamic} proposed a dynamic key-value memory network model for improving the interpretability of the prediction results, and Chen et al.~\cite{chen2018prerequisite} incorporated the knowledge structure information for dealing with the data sparsity problem in knowledge tracing. Experimental results showed that deep models had achieved a great success.

Our work differs from the previous studies as follows. First, existing approaches mainly focus on exploiting students' historical exercising records for their performance prediction, while ignoring the important effects of exercise materials (e.g., knowledge concepts, exercise content). To the best of our knowledge, this work is the first comprehensive attempt that fully explores both student's exercising records and the exercise materials. Second, previous studies follow the common sense that student's next performance only depends on the current states, while our work deeply captures the focused information of students in the history by a novel attention mechanism for improving the prediction. Third, we can well handle the cold-start problem by incorporating exercise correlations without any retraining. Last but not least, our work can achieve good prediction results with interpretability, i.e., we can explain the change of student's knowledge states on explicit knowledge concepts, which is beneficial for many real-world applications, such as explainable exercise recommendation.

\begin{figure}[t]
	\centering
	\includegraphics[scale=0.5]{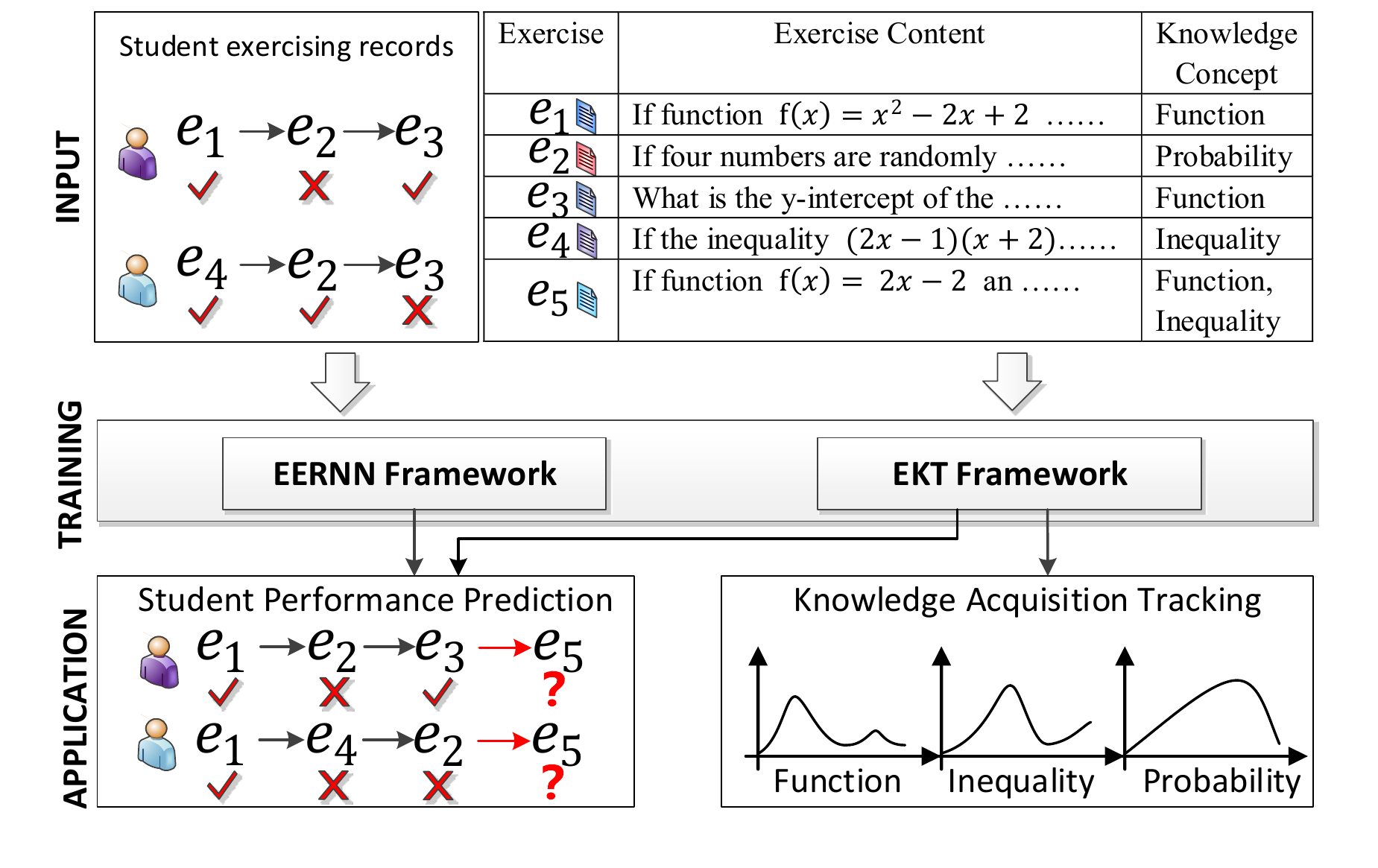}
	\vspace{-0.5cm}
	\caption{An overview of the proposed solution.}
	\vspace{-0.6cm}
	\label{fig:overview}
\end{figure}

\begin{figure*}[ht]
	\centering
	\subfigure[EERNNM with Markov property]{
		\includegraphics[scale=0.43]{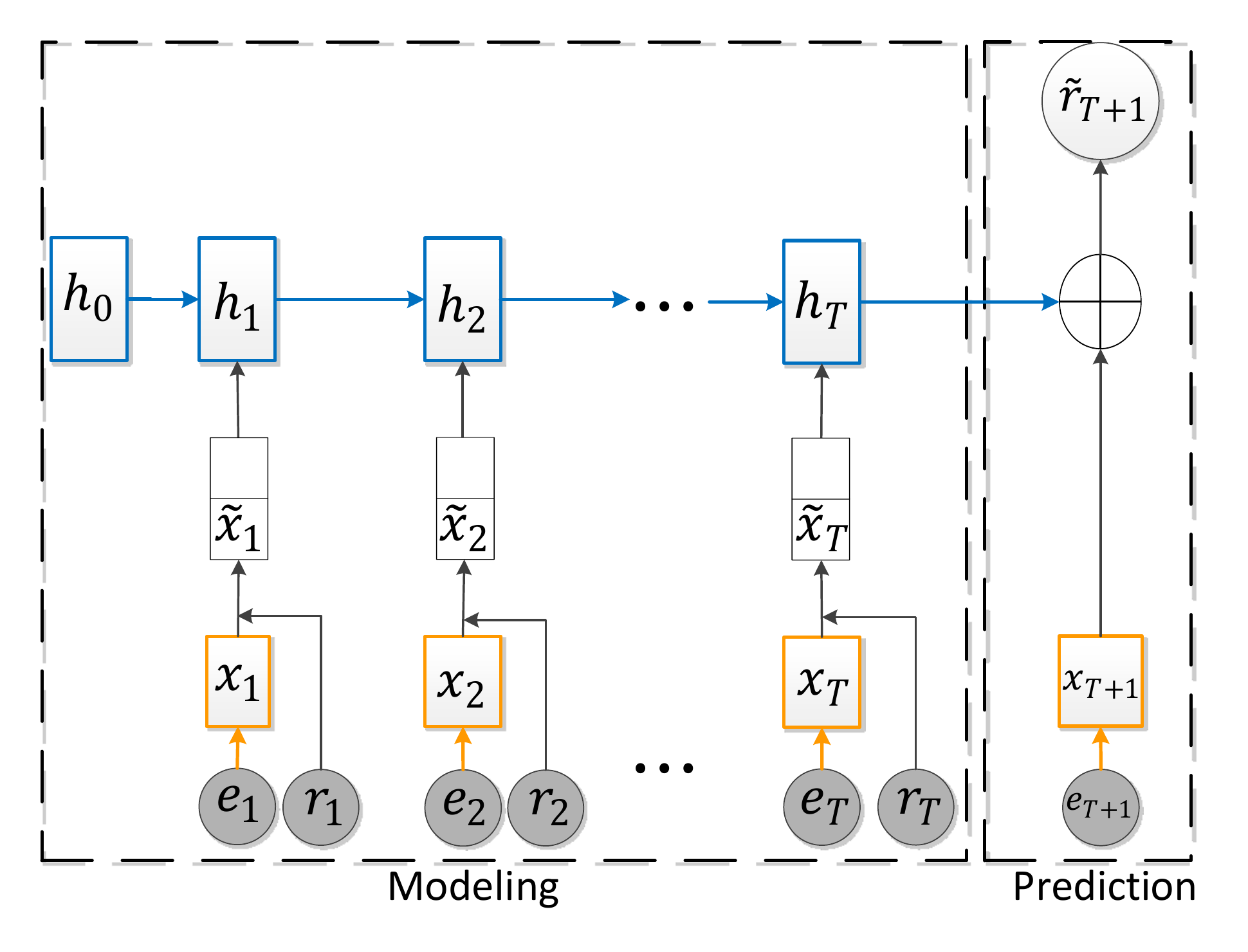}
		\label{fig:subfig:EERNNM}
	}
	\subfigure[EERNNA with Attention mechanism]{
		\includegraphics[scale=0.37]{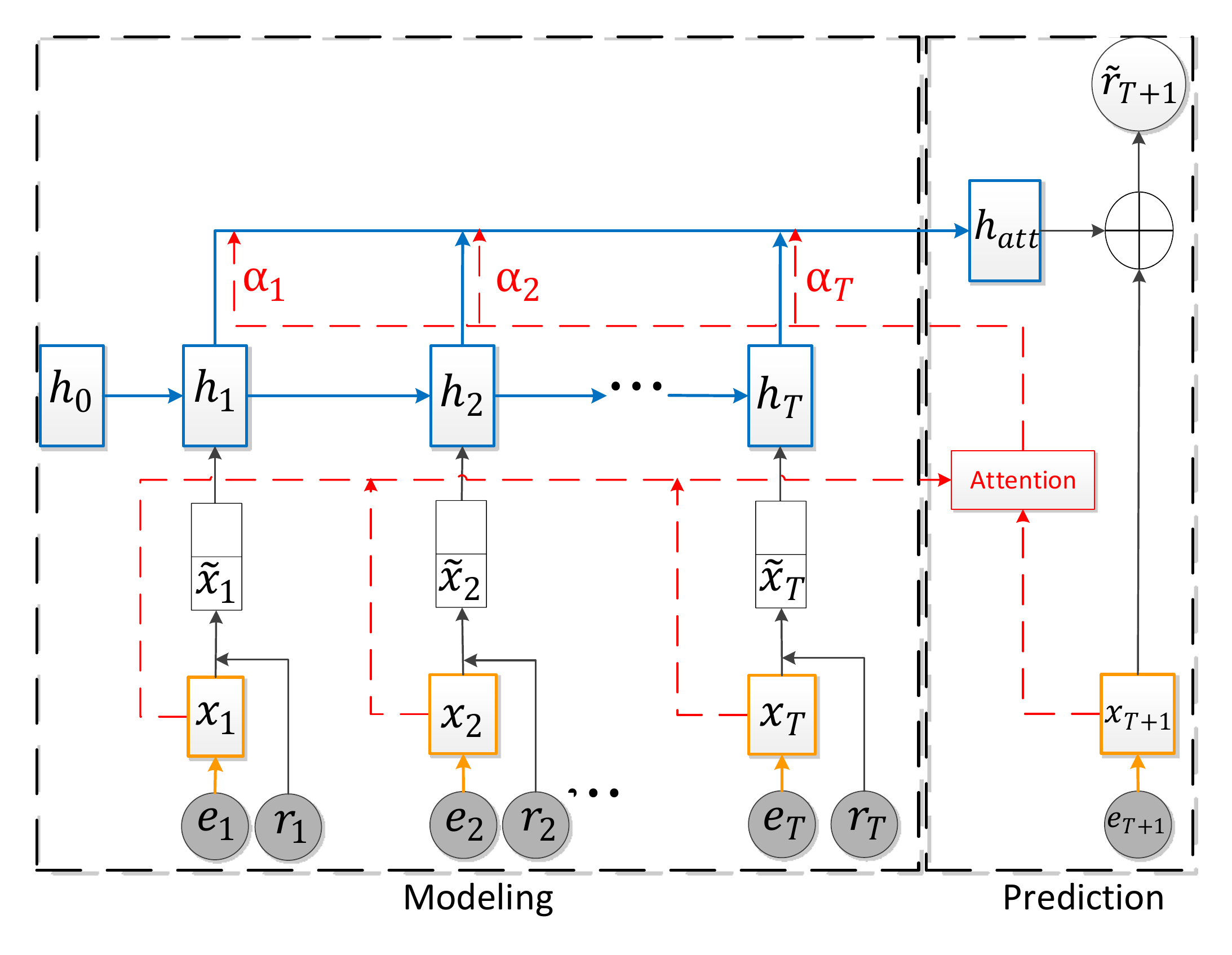}
		\label{fig:subfig:EERNNA}
	}
\vspace{-0.4cm}
	\caption{The architectures of two implementations based on EERNN framework, where the shaded and unshaded symbols denote the observed and latent variables, respectively.}
	\label{fig:EERNN_model_architecture}
	\vspace{-0.6cm}
\end{figure*}

\section{Problem and Solution Overview}
In this section, we first formally define the problem of student performance prediction in intelligent education. Then, we will give the overview of our study.

\textbf{Problem Definition.} In an intelligent education system, suppose there are $|S|$ students and $|E|$ exercises, where students do exercises individually. We record the exercising process of a certain student as $s = \{(e_1, r_1), (e_2, r_2), \dots, (e_T, r_T)\}, s \in S$, where $e_t \in E$ represents the exercise practiced by student $s$ at her exercising step $t$ and $r_t$ denotes the corresponding score. Generally, if student $s$ answers exercise $e_t$ right, $r_t$ equals to 1, otherwise $r_t$ equals to 0. In addition to the logs of student's exercising process, we also consider the materials of exercises (some examples are shown in Fig.~\ref{fig:IntroExample}). Formally, for a certain exercise $e$, we describe it by the text content, which is combined with a word sequence as $e = \{w_1, w_2, \dots, w_M\}$. Also, the exercise $e$ contains its corresponding knowledge concept $k$ coming from all $K$ concepts. Please note that each exercise may contain multiple concepts, e.g., $e_5$ in Fig.~\ref{fig:IntroExample} has two concepts ``Function'' and ``Inequality''. Without loss of generality, in this paper, we represent each student's exercising record as $s = \{(e_1, r_1), (e_2, r_2), \dots, (e_T, r_T)\}$ or $s = \{(k_1, e_1, r_1), (k_2, e_2, r_2), \dots, (k_T, e_T, r_T)\}$, where the former one does not consider the knowledge concept information. Then the problem can be defined as:


\begin{Def} ({\small \textbf{Student Performance Prediction Problem}}).
	Given the exercising logs of each student and the materials of each exercise from exercising step $1$ to $T$, our goal is two-fold: (1) track the change of her knowledge states and estimate how much she masters all $K$ knowledge concepts from step $1$ to $T$; (2) predict the response score $\widetilde{r}_{T+1}$ on the next candidate exercise $e_{T+1}$.
\end{Def}

\textbf{Solution overview.} An overview of the proposed solution is illustrated in Fig.~\ref{fig:overview}. From the figure, given all students' exercising records $S$ with the corresponding exercise materials $E$, we propose a preliminary \emph{E}xercise-\emph{E}nhanced \emph{R}ecurrent \emph{N}eural \emph{N}etwork (EERNN) framework and an improved \emph{E}xercise-aware \emph{K}nowledge \emph{T}racing (EKT) framework. Then, we conduct two applications with the trained models. Specifically, EERNN directly achieves the goal of student performance prediction on future exercises given her sequential exercising records, and EKT is further capable of explicitly tracking the knowledge acquisition of students.

\section{EERNN: Exercise-Enhanced Recurrent Neural Network}
In this section, we first describe the \emph{E}xercise-\emph{E}nhanced \emph{R}ecurrent \emph{N}eural \emph{N}etwork (EERNN) framework that could directly achieve the primary goal of predicting student performance. EERNN is a general framework where we can predict student performance based on different strategies. Specifically, as shown in Fig.~\ref{fig:EERNN_model_architecture}, we propose two implementations under EERNN, i.e., \emph{EERNNM with Markov property} and \emph{EERNNA with Attention mechanism}. Therefore, both models have the same process for modeling student's exercising records yet follow different prediction strategies.

\subsection{Modeling Process of EERNN}
The goal of the modeling process in EERNN framework is to model each student's exercising sequence (with the input notation $s$). From Fig.~\ref{fig:EERNN_model_architecture}, this process contains two main components, i.e., \emph{Exercise Embedding} (marked orange) and \emph{Student Embedding} (marked blue).

\emph{\textbf{Exercise Embedding.}}
As shown in Fig.~\ref{fig:EERNN_model_architecture}, given the exercising process of a student $s = \{(e_1, r_1), (e_2, r_2), \dots, (e_T, r_T)\}$, \emph{Exercise Embedding} learns the semantic representation/encoding $x_i$ of each exercise from its text content $e_i$ automatically.

\begin{figure}[t]
	\centering
	\includegraphics[scale=0.45]{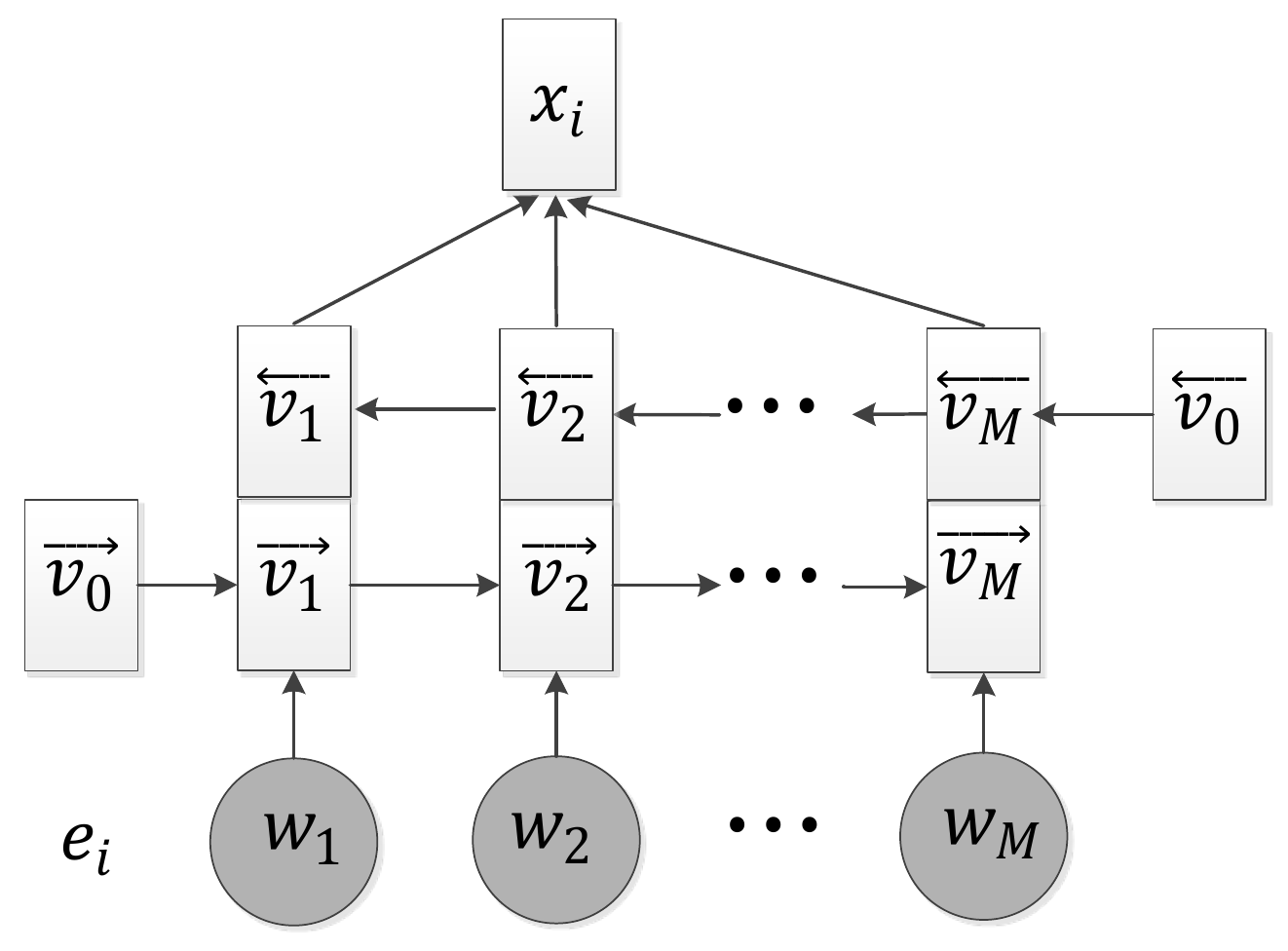}
	\vspace{-0.4cm}
	\caption{Exercise Embedding for exercise $e_i$.}
	\label{fig:ExercisEmb}
	\vspace{-0.6cm}
\end{figure}

Fig.~\ref{fig:ExercisEmb} shows the detailed techniques of \emph{Exercise Embedding}. It is an implementation of a recurrent neural network, which is inspired by the typical one called \emph{Long Short-Term Memory} (LSTM)~\cite{graves2013speech} with minor modifications. Specifically, given the exercise's content with the $M$ words sequence $e_i = \{w_1, w_2, \dots, w_M\}$, we first take \emph{Word2vec}~\cite{mikolov2013distributed} to transform each word $w_i$ in exercise $e_i$ into a $d_0$-dimensional pre-trained word embedding vector. After the initialization, \emph{Exercise Embedding} updates the hidden state $v_m \in \mathbb{R}^{d_{v}}$ of each word $w_m$ at the $m$-th word step with the previous hidden state $v_{m-1}$ in a formula as:
\begin{small}
	\begin{eqnarray}
	\label{equ:ExerEmb}
	&&i_m = \sigma(\mathbf{Z_{wi}^E} w_m + \mathbf{Z_{vi}^E} v_{m-1} + \mathbf{b_i^E}), \nonumber \\
	&&f_m = \sigma(\mathbf{Z_{wf}^E} w_m + \mathbf{Z_{vf}^E} v_{m-1} + \mathbf{b_f^E}), \nonumber \\
	&&o_m = \sigma(\mathbf{Z_{wo}^E} w_m + \mathbf{Z_{vo}^E} v_{m-1} + \mathbf{b_o^E}), \\
	&&c_m = f_m \cdot c_{m-1} + i_m \cdot \tanh(\mathbf{Z_{wc}^E} w_m + \mathbf{Z_{vc}^E} v_{m-1} + \mathbf{b_c^E}), \nonumber \\
	&&v_m = o_m \cdot \tanh(c_m), \nonumber
	\end{eqnarray}
\end{small}
\noindent where $i_m, f_m, o_m$ represent the three gates, i.e., input, forget, output, respectively. $c_m$ is a cell memory vector. $\sigma(x)$ is the non-linear \emph{sigmoid} activation function and $\cdot$ denotes the element-wise product between vectors. Besides, the input weighted matrices $\mathbf{Z_{w*}^E} \in \mathbb{R}^{d_v \times d_0}$, recurrent weighted matrices $\mathbf{Z_{v*}^E} \in \mathbb{R}^{d_v \times d_v}$ and bias weighted vectors $\mathbf{b_*^E} \in \mathbb{R}^{d_v}$ are all the network parameters in \emph{Exercise Embedding}.

Traditional LSTM model learns each word representation by a single direction network and can not utilize the contextual texts from the future word token~\cite{tan2015lstm}. To make full use of the contextual word information of each exercise, we build a bidirectional LSTM considering the word sequence in both forward and backward directions. As illustrated in Fig.~\ref{fig:ExercisEmb}, at each word step $m$, the forward layer with hidden word state $\overrightarrow{v}_{m}$ is computed based on both the previous hidden state $\overrightarrow{v}_{m-1}$ and the current word $w_m$; while the backward layer updates hidden word state $\overleftarrow{v}_{m}$ with the future hidden state $\overleftarrow{v}_{m+1}$ and the current word $w_m$. As a result, each word's hidden representation $v_m$ can be calculated with the concatenation of the forward state and backward state as $v_m = concatenate(\overrightarrow{v}_{m}, \overleftarrow{v}_{m})$.

After that, to obtain the whole semantic representation of exercise $e_i$, we exploit the element-wise max pooling operation to merge $M$ words' contextual representations into a global embedding $x_i \in \mathbb{R}^{2d_v}$ as $x_i = \max(v_1, v_2, \dots, v_M)$.

It is worth mentioning that \emph{Exercise Embedding} directly learns the semantic representation of each exercise from its text without any expert encoding. It can also automatically capture the characteristics (e.g., difficulty) of exercises and distinguish their individual differences.

\emph{\textbf{Student Embedding.}}
After obtaining each exercise representation $x_i$ from the text content $e_i$ by \emph{Exercise Embedding}, \emph{Student Embedding} aims at modeling the whole exercising process of students and learning the hidden representations of students, which we called \emph{student states}, at different exercising steps combined with the influence of student performance in the history. As shown in Fig.~\ref{fig:EERNN_model_architecture}, EERNN assumes that the student states are influenced by both the exercises and the corresponding scores she got.

Along this line, we exploit a recurrent neural network for \emph{Student Embedding} with the input of a certain student's exercising process $s = \{(x_1, r_1), (x_2, r_2), \dots, (x_T, r_T)\}$. Specifically, at each exercising step $t$, the input to the network is a combined encoding with both exercise embedding $x_t$ and the corresponding score $r_t$. Since students getting right response (i.e., score 1) and wrong response (i.e., score 0) to the same exercise actually reflect their different states, we need to find an appropriate way to distinguish these different effects for a specific student.


Methodology-wise, we first extend the score value $r_t$ to a feature vector $\mathbf{0} = (0, 0, \dots, 0)$ with the same $2d_v$ dimensions of exercise embedding $x_t$ and then learn the combined input vector $\widetilde{x}_t \in \mathbb{R}^{4d_v}$ as:
\begin{equation}
	\label{equ:exerAndScore}
	\widetilde{x}_t=
		\begin{cases}
			[x_t \oplus \mathbf{0}] & \text{if}\quad r_t = 1, \\
			[\mathbf{0} \oplus x_t] & \text{if}\quad r_t = 0, \\
		\end{cases}
\end{equation}
\noindent where $\oplus$ is the operation that concatenates two vectors.

With the combined exercising sequence of a student $s = \{\widetilde{x}_1, \widetilde{x}_2, \dots, \widetilde{x}_T\}$, the hidden student state $h_t \in \mathbb{R}^{d_h}$ at her exercising step $t$ is updated based on the current input $\widetilde{x}_t$ and the previous state $h_{t-1}$ in a recurrent formula as:
\begin{equation}
	\label{eq:rnn}
	h_t = RNN(\widetilde{x}_t, h_{t-1}; \theta_h).
\end{equation}

In the literature, there are many variants of the RNN forms~\cite{graves2013speech,chung2014empirical}. In this paper, considering the fact that the length of student's exercising sequence can be long, we also implement Eq.~(\ref{eq:rnn}) by the sophisticated LSTM form, i.e., $h_t = LSTM(\widetilde{x}_t, h_{t-1};\theta_h)$, which could preserve more long-term dependency in the sequence as:
\begin{small}
	\begin{eqnarray} \label{equ:LearnEmb}
	&&i_t = \sigma(\mathbf{Z_{\widetilde{x}i}^S} \widetilde{x}_t + \mathbf{Z_{hi}^S} h_{t-1} + \mathbf{b_i^S}), \nonumber \\
	&&f_t = \sigma(\mathbf{Z_{\widetilde{x}f}^S} \widetilde{x}_t + \mathbf{Z_{hf}^S} h_{t-1} + \mathbf{b_f^S}), \nonumber \\
	&&o_t = \sigma(\mathbf{Z_{\widetilde{x}o}^S} \widetilde{x}_t + \mathbf{Z_{ho}^S} h_{t-1} + \mathbf{b_o^S}), \\
	&&c_t = f_t \cdot c_{t-1} + i_t \cdot \tanh(\mathbf{Z_{\widetilde{x}c}^S} \widetilde{x}_t + \mathbf{Z_{hc}^S} h_{t-1} + \mathbf{b_c^S}), \nonumber \\
	&&h_t = o_t \cdot \tanh(c_t), \nonumber
	\end{eqnarray}
\end{small}
\noindent where $\mathbf{Z_{\widetilde{x}*}^S} \in \mathbb{R}^{d_h \times 4d_v}, \mathbf{Z_{h*}^S} \in \mathbb{R}^{d_h \times d_h}$ and $\mathbf{b_*^S} \in \mathbb{R}^{d_h}$ are the parameters in \emph{Student Embedding}.

Particularly, the input weight matrix  $\mathbf{Z_{\widetilde{x}*}^S} \in \mathbb{R}^{d_h \times 4d_v}$ in Eq.~(\ref{equ:LearnEmb}) can be divided into two parts, i.e., the positive one $\mathbf{Z_{\widetilde{x}*}^{S+}} \in \mathbb{R}^{d_h \times 2d_v}$ and the negative one $\mathbf{Z_{\widetilde{x}*}^{S-}} \in \mathbb{R}^{d_h \times 2d_v}$, which can separately capture the influences of exercise $e_i$ with both right and wrong responses for a specific student during her exercising process. Based on these two types of parameters, \emph{Student Embedding} can naturally model the exercising process to obtain student states by integrating both the exercise contents and the response scores.

\subsection{Prediction Output of EERNN}
After modeling the exercising process of each student from exercising step 1 to $T$, we now introduce the detailed strategies of predicting her performance on exercise $e_{T+1}$. Psychological results have claimed that student-exercise performances depend on both the student states and the exercise characteristics~\cite{dibello200631a}. Following this finding, we propose two implementations of prediction strategies under EERNN framework, i.e., a straightforward yet effective \emph{EERNNM with Markov property} and a more sophisticated \emph{EERNNA with Attention mechanism}, based on both the learned student states $\{h_1, h_2, \dots, h_T\}$ and the exercise embeddings $\{x_1, x_2, \dots, x_T\}$.

\emph{\textbf{EERNN with Markov Property.}}
For a typical sequential prediction task, Markov property is a well understood and widely used theory which assumes that the next state depends only on the current state and not on the sequences that precede it~\cite{rabiner1986introduction}. Given this theory, as shown in Fig.~\ref{fig:subfig:EERNNM}, when an exercise $e_{T+1}$ at step $T+1$ is posted to a student, EERNNM (1) assumes that the student applies current state $h_T$ to solve the exercise; (2) leverages \emph{Exercise Embedding} to extract the semantic representation $x_{T+1}$ from exercise text $e_{T+1}$; (3) predicts her performance $\widetilde{r}_{T+1}$ on exercise $e_{T+1}$ as following formulas:
\begin{eqnarray}
	\label{equ:EERNNM}
	&& y_{T+1} = ReLU (\mathbf{W_1} \cdot [h_T \oplus x_{T+1}] + \mathbf{b_1}), \nonumber  \\
	&& \widetilde{r}_{T+1} = \sigma (\mathbf{W_2} \cdot y_{T+1} + \mathbf{b_2}),
\end{eqnarray}
\noindent where $y_{T+1} \in \mathbb{R}^{d_y}$ denotes the overall presentation for prediction at ($T+1$)-th exercising step. \{$\mathbf{W_1}, \mathbf{W_2}, \mathbf{b_1}, \mathbf{b_2}$\} are the parameters. $\sigma(x)$ is the \emph{Sigmoid} activation function $\sigma(x)=\frac{1}{1+\exp(-x)}$ and $\oplus$ is the concatenation operation.

EERNNM presents a straightforward yet effective way for student performance prediction. However, in most cases, since the current student state $h_T$ is the last hidden state of the LSTM-based architecture in \emph{Student Embedding}, it may discard some important information when the sequence is too long, which is called the \emph{Vanish} problem~\cite{hochreiter1997long}. Thus, the learned student state by EERNNM may be somewhat unsatisfactory for future performance prediction. To address this question, we further propose another sophisticated prediction strategy, i.e., \emph{EERNNA with Attention mechanism}, to enhance the effects of important student states in the exercising process for prediction.

\emph{\textbf{EERNNA with Attention Mechanism.}}
In Fig.~\ref{fig:IntroExample}, students may get similar scores on similar exercises, e.g., student $s_1$ answers the exercises $e_1$ and $e_3$ right due to the possible reason that the both exercises are similar because of the same knowledge concept ``Function''.

According to this observation, as the red lines illustrated in Fig.~\ref{fig:subfig:EERNNA}, EERNNA assumes that the student state at ($T+1$)-th exercising step is a weighted sum aggregation of all historical student states based on the correlations between exercise $e_{T+1}$ and the historical ones $\{e_1, e_2, \dots, e_T\}$. Formally, at next step $T+1$, we define the attentive state vector $h_{att}$ of student as:
\begin{equation}
	 \label{equ:hatt}
	\centering
	h_{att} = \sum_{j=1}^{T} \alpha_j h_{j}, \  \alpha_j = cos(x_{T+1}, x_{j}),
\end{equation}
\noindent where $x_j$ is the exercise embedding at $j$-th exercising step and $h_j$ is the corresponding student state in the history. \emph{Cosine Similarities} $\alpha_j$ are denoted as the attention scores for measuring the importance of each exercise $e_j$ in the history for new exercise $e_{T+1}$.

After obtaining attentive student state at step $T+1$, EERNNA predicts the performance of this student on exercise $e_{T+1}$ with the similar operation in Eq.~(\ref{equ:EERNNM}) by replacing $h_T$ with $h_{att}$.

Particularly, through \emph{Exercise Embedding}, our attention scores $\alpha_j$ not only measure the similarity between exercises from syntactic perspective but also capture the correlations from semantic view (e.g., difficulty correlation), benefiting student state representation for student performance prediction and model explanation. We will conduct experimental analysis for this attention mechanism.

\subsection{Model Learning}
The whole parameters to be updated in both proposed models mainly come from three parts, i.e., parameters in \emph{Exercise Embedding} $\{\mathbf{Z_{w*}^E}, \mathbf{Z_{v*}^E}, \mathbf{b_*^E}\}$, parameters in \emph{Student Embedding} $\{\mathbf{Z_{\widetilde{x}*}^S}, \mathbf{Z_{h*}^S}, \mathbf{b_*^S}\}$ and parameters in \emph{Prediction Output} $\{\mathbf{W_*}, \mathbf{b_*}\}$. The objective function of EERNN is the negative log likelihood of the observed sequence of student's exercising process from step 1 to $T$. Formally, at $t$-th step, let $\widetilde{r}_t$ be the predicted score on exercise $e_t$ through EERNN framework, $r_t$ is the actual binary score, thus the overall loss for a certain student is defined as:
\begin{equation}
	\label{equ:loss}
	\mathcal{L} = - \sum_{t=1}^{T} (r_t \log \widetilde{r}_t + (1 - r_t)\log( 1 - \widetilde{r}_t)).
\end{equation}

The objective function is minimized by the Adam optimization~\cite{kingma2014adam}. Details will be specified in the experiments.

\begin{figure*}[ht]
	\centering
	\subfigure[EKTM with Markov property]{
		\includegraphics[scale=0.275]{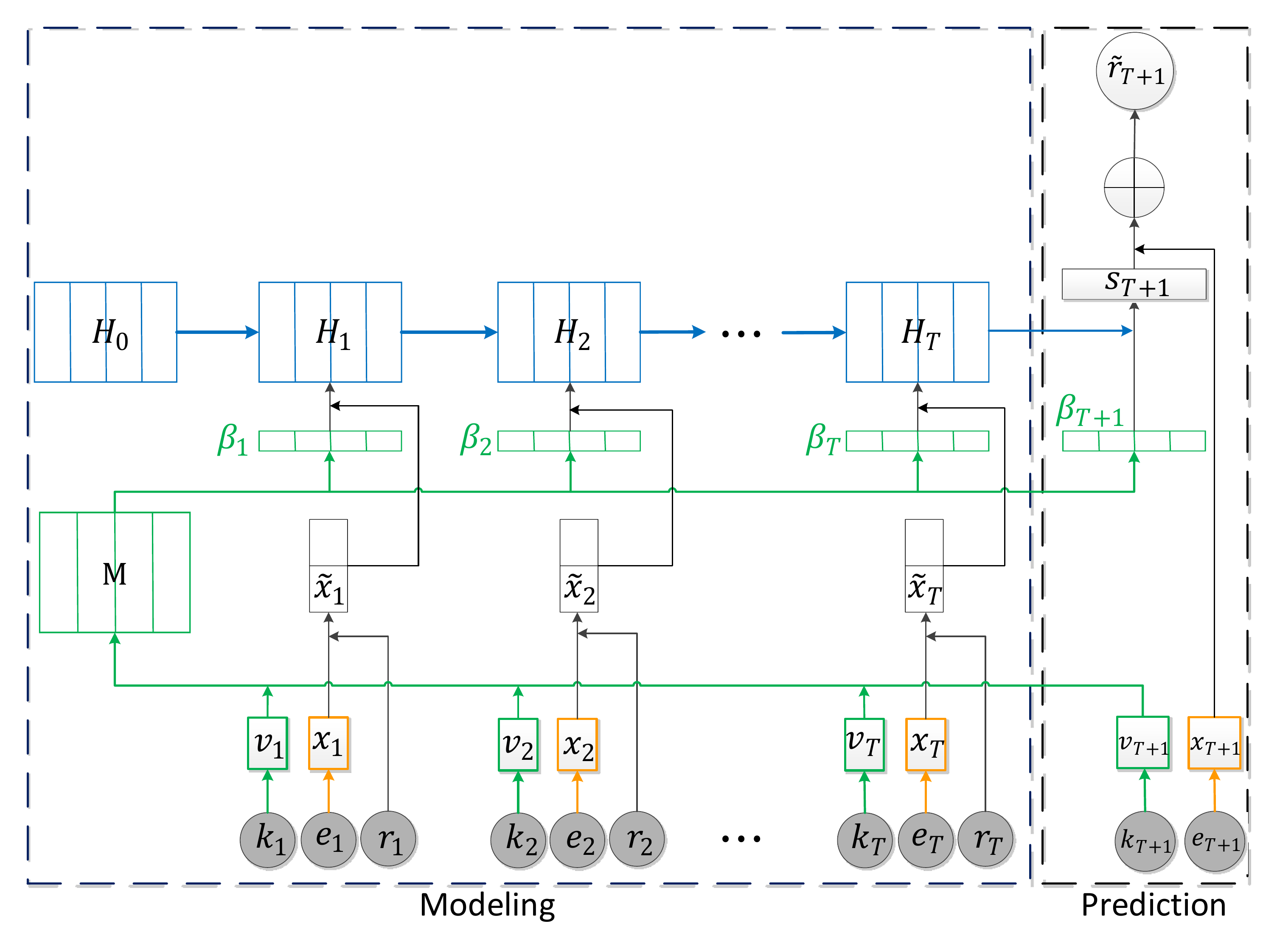}
		\label{fig:subfig:EKTM}
	}
	\subfigure[EKTA with Attention mechanism]{
		\includegraphics[scale=0.24]{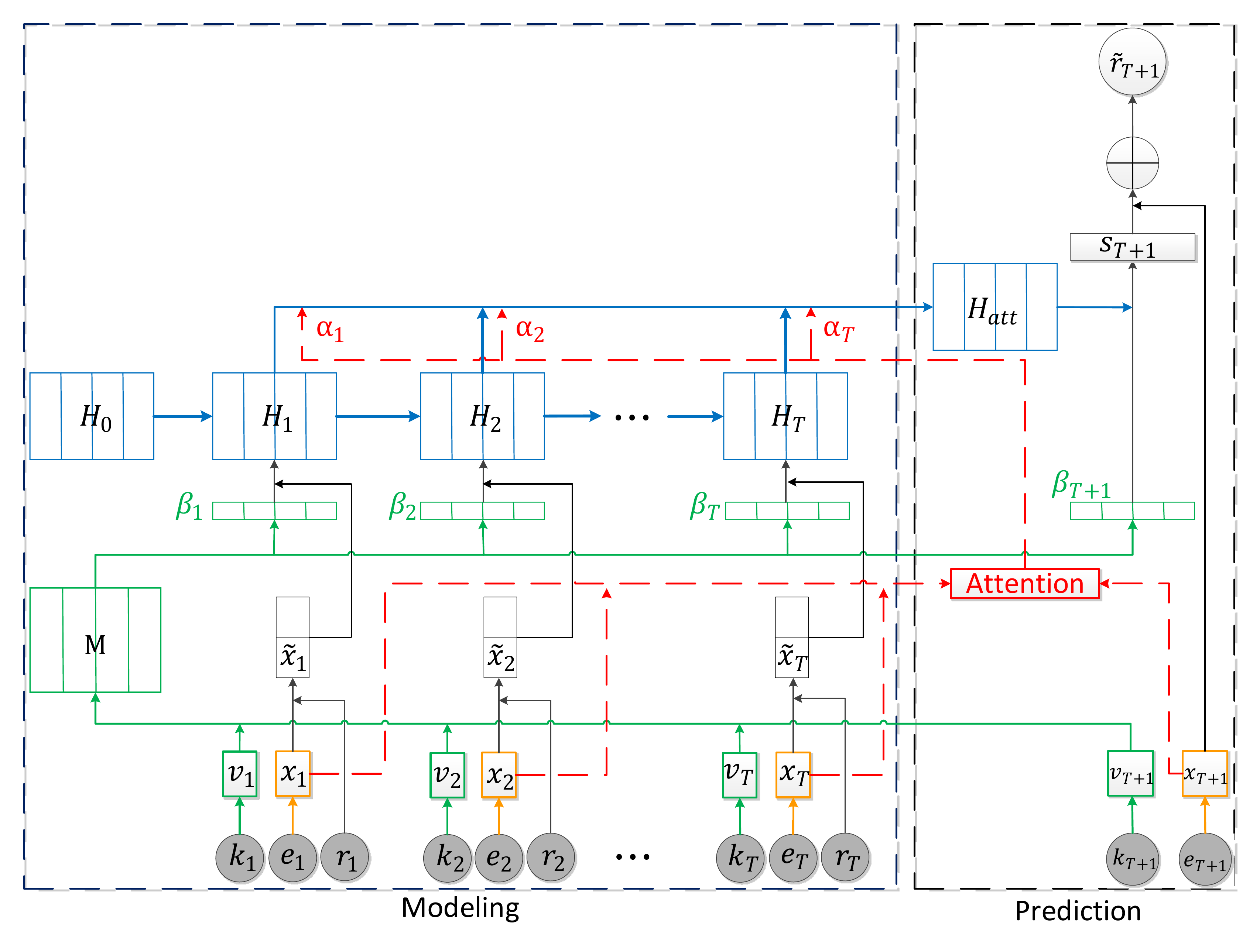}
		\label{fig:subfig:EKTA}
	}
\vspace{-0.4cm}
	\caption{The architectures of two implementations based on EKT framework, where the shaded and unshaded symbols denotes the observed and latent variables, respectively.}
	\label{fig:EKT_model_architecture}
	\vspace{-0.6cm}
\end{figure*}

\section{EKT: Exercise-aware Knowledge Tracing} \label{sec:subsec:EKT}
EERNN can effectively deal with the problem of predicting student performance on future exercises. However, during the modeling, we just summarize and track a student's knowledge states on all concepts in one integrated hidden vector (i.e., $h_t$ in Eq.~(\ref{equ:LearnEmb})), and this is sometimes unsatisfied because it is hard to explicitly explain how much she has mastered a certain knowledge concept (e.g., ``Function''). In fact, during the exercising process of a certain student, when an exercise is given, she usually applies her relevant knowledge to solve it. Correspondingly, her performance on the exercise, i.e., whether or not she answers it right, can also reflect how much she has mastered the knowledge~\cite{corbett1994knowledge,zhang2017dynamic}. For example, we could conclude that the student in Fig.~\ref{fig:IntroExample} has well mastered the ``Function'' and ``Inequality'' concepts but needs to devote more energy to the less familiar one ``Probability''. Thus, it is valuable if we could remind her about this finding so that she could prepare the target training about ``Probability'' herself. Based on the above understanding, in this section, we further address the problem of tracking student's knowledge acquisition on multiple explicit concepts. We extend the current EERNN and propose an explainable \emph{E}xercise-aware \emph{K}nowledge \emph{T}racing (EKT) framework by incorporating the information of knowledge concepts existed in each exercise.

Specifically, we extend the knowledge states of a certain student from the integrated vectorial representation in EERNN, i.e., $h_t \in \mathbb{R}^{d_h}$, to a matrix with multiple vectors, i.e., $H_t \in \mathbb{R}^{d_h \times K}$, where each vector represents how much she has mastered an explicit knowledge concept (e.g., ``Function''). Meanwhile, in EKT, we assume the student's knowledge state matrix $H_t$ changes over time influenced by both text content (i.e., $e_t$) and knowledge concept (i.e., $k_t$) of each exercise. Fig.~\ref{fig:EKT_model_architecture} illustrates the overall architecture of EKT. Comparing it with EERNN (Fig.~\ref{fig:EERNN_model_architecture}), besides the \emph{Exercise Embedding} module, another module (marked green), which we called \emph{Knowledge Embedding}, is incorporated in the modeling process. With this additional facility, we can naturally extend the proposed prediction strategies EERNNM and EERNNA to \emph{EKTM with Markov property} and \emph{EKTA with Attention mechanism}, respectively. In the following, we first introduce the way to implement the \emph{Knowledge Embedding} module, followed by the details of EKTM and EKTA.

\emph{\textbf{Knowledge Embedding.}}
Given the student's exercising process $s = \{(k_1, e_1, r_1), (k_2, e_2, r_2), \dots, (k_T, e_T, r_T)\}$, the goal of \emph{Knowledge Embedding} is to explore the impacts of each exercise on improving student states from this exercise's knowledge concepts $k_t$, and this impact weight is denoted by $\beta_t$. Intuitively, at step $t$, if this exercise is related to the $i$-th concept, we can just consider the impact of this specific concept without others' influences, i.e., $\beta_t^j = 1$ if $j = i$, otherwise $\beta_t^j =0$, $0 \leq i,j \leq K$. However, in educational psychology, some findings indicate that the knowledge concepts in one specific domain (e.g., mathematics) are not isolated but contain correlations with each other~\cite{zhang2017dynamic}. Hence, in our modeling, we assume that learning one concept, for a certain student, could also affect the acquisition of other concepts. Thus, it is necessary to quantify these correlation weights among all $K$ concepts in the knowledge space.

Along this line, as the module (marked in green) shown in Fig.~\ref{fig:EKT_model_architecture}, we investigate and propose a static memory network for calculating knowledge impact $\beta_t$. Specifically, it is inspired by the memory-augmented neural network~\cite{graves2016hybrid,sukhbaatar2015end}, which has been successfully adopted in many applications, such as question answering~\cite{xiong2016dynamic}, language modeling~\cite{kumar2016ask} and one-shot learning~\cite{santoro2016meta}. It usually contains an external memory component that can store the stable information. Then, during the sequence, it can read each input and write the storage information from the memory for influencing its long-term dependency. Considering this property, we set up a memory module with a matrix $\mathbf{M} \in \mathbb{R}^{d_k \times K}$ to store the representations of $K$ knowledge  concepts by $d_k$-dimensional features.


Mathematically, at each exercising step $t$, when an exercise $e_t$ comes, we first set its knowledge concept to be a one-hot encoding $k_t \in {\{0,1\}}^K$ with the dimension equaling to the total number $K$ of all concepts. Since the intuitive one-hot representation is too sparse for modeling~\cite{goldberg2014word2vec}, we utilize an embedding matrix $\mathbf{W_k} \in \mathbb{R}^{K \times d_k}$ to transfer the initial knowledge encoding $k_t$ into a low-dimensional vector $v_t \in \mathbb{R}^{d_k}$ with continuous values as: $v_t = \mathbf{W_k}^{\rm T} k_t$.

 After that, the impact weight $\beta_t^i (1 \le i \le K)$ on the $i$-th concept from exercise $e_t$'s knowledge concept $k_t$ is further calculated by the softmax operation of the inner product between the given concept encoding $v_t$ and each knowledge memory vector in the memory module $\mathbf{M}_i$ as:
\begin{equation}
\label{eq:knowledge_weight}
\beta_t^i =  \mathrm{Softmax}(v_t^{\rm T} \mathbf{M}_{i}) = \frac{\exp(v_t^{\rm T} \mathbf{M}_i)}{\sum_{i=1}^{K}(\exp(v_t^{\rm T} \mathbf{M}_i))}.
\end{equation}

\emph{\textbf{Student Embedding.}}
With the knowledge impact $\beta_t$ of each exercise, an improved \emph{Student Embedding} will further specify each knowledge acquisition of a certain student during her exercising process. Thus, EKT could naturally track student's knowledge states on multiple concepts simultaneously, benefiting the interpretability.

Methodology-wise, at the exercising step $t$, we also update one of a student's specific knowledge state $H_t^i \in \mathbb{R}^{d_h} (1 \le i \le K)$ by the LSTM network after she answers the exercise $e_t$: 
\begin{equation}
	\label{eq:lstm_H}
	H_t^i = LSTM(\widetilde{x}_t^i, H_{t-1}^i; \theta_{H^i}),
\end{equation}
here we replace the original input $\widetilde{x}_t$ with a new joint one $\widetilde{x}_t^i$ which is computed in the formula as: $\widetilde{x}_t^i = \beta_t^i  \widetilde{x}_t$, where $\widetilde{x}_t$ is the same encoding that combines the effects of both the exercise $e_t$ she practices and the score $r_t$ she gets (Eq.~(\ref{equ:exerAndScore})).

After modeling student's historical exercising process, in the prediction part of EKT, the performance of each student is predicted based on three types of factors, i.e., her historical knowledge states $\{H_1, H_2, \cdots, H_T\}$, the embeddings of the exercises she practiced $\{x_1, x_2, \cdots, x_T\}$, and the materials $k_{T+1}$ and $e_{T+1}$ of the candidate exercise.

\emph{\textbf{EKTM with Markov property.}}
Similar to EERNNM, EKTM follows the straightforward Markov property that assumes student performance on further exercise only depends on her current knowledge state $H_T$. Specifically, as shown in Fig.~\ref{fig:subfig:EKTM}, when the exercise $e_{T+1}$ is posted, EKTM first integrates student's mastery on this exercise with its knowledge impacts $\beta_{T+1}$ as:
\begin{equation}
	\label{eq:student_level}
	s_{T+1} = \sum_{i=1}^{K} \beta_{T+1}^i H_{T}^i,
\end{equation}
then predicts her performance $\widetilde{x}_{T+1}$ by changing the similar operation in Eq.~(\ref{equ:EERNNM}) as:
\begin{eqnarray}
	\label{eq:EKTM}
	&& y_{T+1} = ReLU (\mathbf{W_3} \cdot [s_{T+1} \oplus x_{T+1}] + \mathbf{b_3}), \nonumber  \\
	&& \widetilde{r}_{T+1} = \sigma (\mathbf{W_4} \cdot y_{T+1} + \mathbf{b_4}),
\end{eqnarray}
where \{$\mathbf{W_3}, \mathbf{W_4}, \mathbf{b_3}, \mathbf{b_4}$\} are the parameters.

\emph{\textbf{EKTA with Attention mechanism.}}
EKTA also follows the sophisticated Attention mechanism to enhance the effect of important states in the history for predicting student's future performance, which is shown in Fig.~\ref{fig:subfig:EKTA}. Here, a small modification compared with EERNNA is that we extend the attentive state vector $h_{att}$ of student (Eq.~(\ref{equ:hatt})) to a matrix one $H_{att}$, where each knowledge state slot $H_{att}^i (1 \le i \le K)$ can be computed as:
\begin{equation}
	\label{equ:Hatt}
	\centering
	H_{att}^i = \sum_{j=1}^{T} \alpha_j H_{j}^i, \  \alpha_j = cos(x_{T+1}, x_{j}).
\end{equation}
Then, EKTA generates the prediction on exercise $e_{T+1}$ with Eq.~(\ref{eq:student_level}) and Eq.~(\ref{eq:EKTM}) by replacing $H_T$ with $H_{att}$.

After that, we can train EKT by minimizing the same objective function in Eq.~(\ref{equ:loss}). Please note that during our modeling, EKT framework could enhance the interpretability of the learned matrix $H_t$ through the impact weight $\beta_t$, which could tell us the mastery levels on each concept of a certain student at exercising step $t$, and we will discuss the details in the next section.

\section{Application}
After discussing the training stage of both EERNN and EKT, we now present the way to apply EERNN and EKT based models to achieve two motivating goals, i.e., student performance prediction and knowledge acquisition tracking.

\textbf{Student Performance Prediction.} As one of the primary applications in intelligent education, student performance prediction helps provide better proactive services to students, such as personalized exercise recommendation~\cite{kuh2011piecing}. Both EERNN and EKT can directly achieve this goal.

Specifically, with the trained EERNN (EKT) model $\mathcal{M}$, given an individual student and her exercising record $s^{p} = \{(k_1^p, e_1^p, r_1^p), (k_2^p, e_2^p, r_2^p), \dots, (k_T^p, e_T^p, r_T^p)\}$, we could predict her performance on the next exercise $e_{T+1}^p$ by the following steps: (1) apply model $\mathcal{M}$ to fit her exercising process $s^{p}$ to get the student state at step $T$ for prediction (i.e., $h_T^p$ in EERNNM or $H_{T}^p$ in EKTM); (2) extract exercise representation $x_{T+1}^p$ and knowledge impact $\beta_{T+1}$ by \emph{Exercise Embedding} and \emph{Knowledge Embedding}; (3) predict her performance $\widetilde{r}_{T+1}^p$ with Eq.~(\ref{equ:EERNNM}) (Eq.~(\ref{eq:EKTM})). Similarly, EERNNA (EKTA) generates the prediction by replacing $h_T^p$ ($H_T^p$) with $h_{att}^p$ ($H_{att}^p$).

Please note that student $s^{p}$ can be either any one that exists in the training stage or a new student that never shows up. Equally, exercise $e_i^p$ in $s^{p}$ can also be either a learned exercise or any new exercise. Specifically, when a new student without any historical record is coming, at step 1, EERNN (EKT) can model her first state $h_1$ ($H_1$) and make performance prediction by the non-personalized prior $h_0$  in Fig.~\ref{fig:EERNN_model_architecture} ($H_0$ in Fig.~\ref{fig:EKT_model_architecture}), i.e., the state generated from all trained student records. After that, EERNN (EKT) can fit her own exercising process and make personalized predictions on the following exercises. Similarly, when a new exercise is coming, Exercise Embedding (Knowledge Embedding) in EERNN (EKT) can learn its representation (impact) only based on its original content (concept). Last but not least, all the prediction part of EERNN (EKT) do not require any model retraining. Therefore, EERNN (EKT) can naturally deal with the cold-start problem when making predictions for new students and new exercises.

\textbf{Knowledge Acquisition Tracking.} It is of great importance to remind students about how much they have mastered each knowledge concept (e.g., with the mastery level ranges from 0 to 1) as they can be motived to conduct the target training in time for practicing more efficiently~\cite{grossman2011transfer}. As mentioned earlier, the EKT framework has a good ability to track student's knowledge acquisition with the learned states $\{H_1, H_2, \cdots, H_T\}$. Inspired by~\cite{zhang2017dynamic}, we introduce the way to estimate the knowledge mastery level of students.

\begin{figure}[tp]
	\centering
	\includegraphics[scale=0.45]{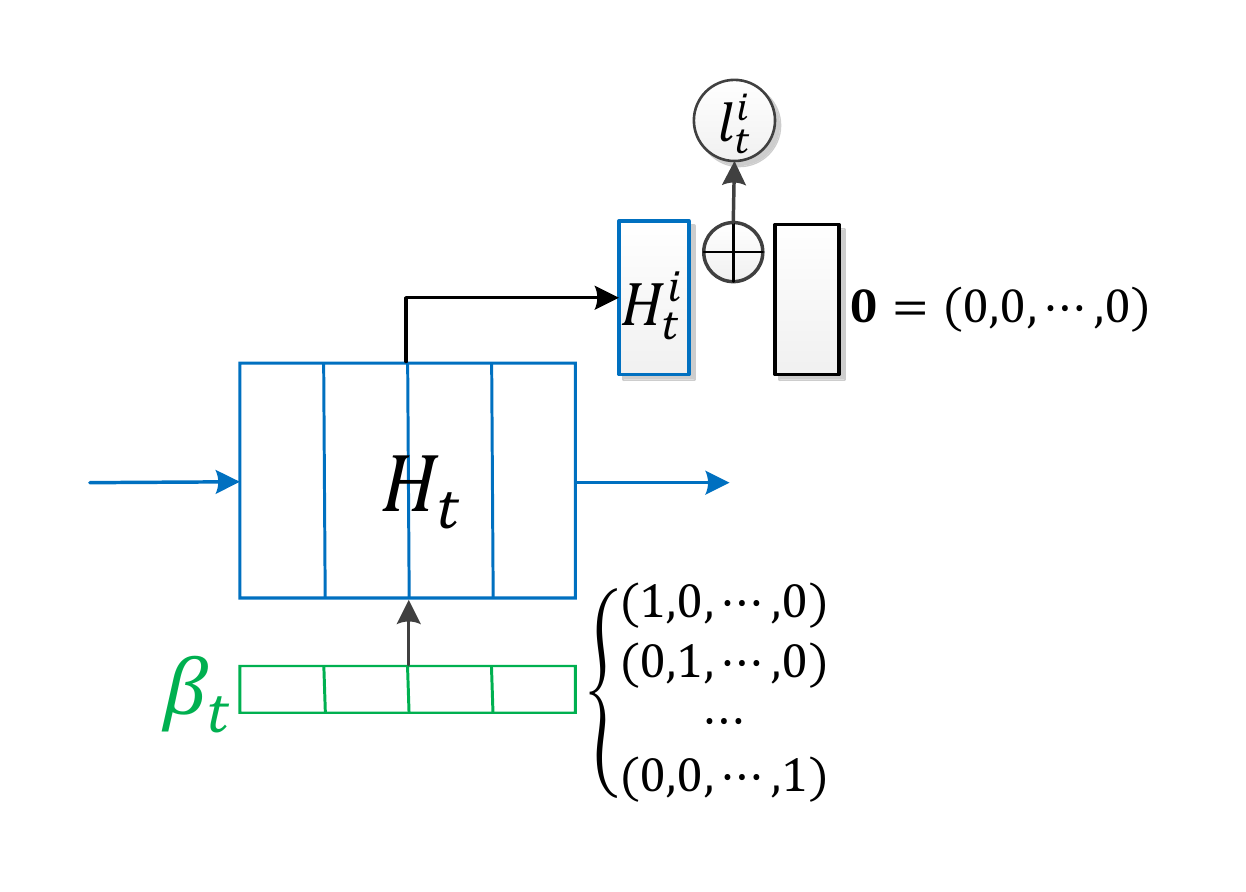}
	\vspace{-0.4cm}
	\caption{The process of mastery level estimation on knowledge concepts.}
	\vspace{-0.3cm}
	\label{fig:mastery_process}
\end{figure}

In the prediction part, at each step $t$, please note that Eq.~(\ref{eq:EKTM}) predicts student performance on a specific exercise $e_t$ from two kinds of inputs: the student's integrated mastery for this exercise (i.e., $s_t$) and the individual exercise embedding (i.e., $x_t$). Thus, if we just want to estimate her mastery of the $i$- th specific concept without any exercise input, we can change $s_t$ by her state in $H_t$ on this concept (i.e., $H_t^i$), and meanwhile, omit the input exercise embedding $x_t$. Fig.~\ref{fig:mastery_process} shows the detailed process of this mastery level estimation on knowledge concepts. Specifically, given a student's exercising record $s = \{(k_1, e_1, r_1), (k_2, e_2, r_2), \dots, (k_T, e_T, r_T)\}$, we first obtain her knowledge state $H_t$ at step $t$ by fitting the record from $1$ to $t$ with the trained EKT. Then, to estimate her mastery of the $i$-th specific concept, we construct the impact weight $\beta_t = (0, \cdots, 1, \cdots, 0)$, where the value in $i$-th dimension equals to 1, and also extract her knowledge state $H_t^i$ on $i$-th concept by Eq.~(\ref{eq:student_level}). After that, we can change Eq.~(\ref{eq:EKTM}) and finally estimate her mastery level $l_t^i$ by:
\begin{eqnarray}
	\label{eq:mastery}
	&& y_t^i = ReLU (\mathbf{W_3} \cdot [H_t^i \oplus \mathbf{0}] + \mathbf{b_3}), \nonumber  \\
	&& l_t^i = \sigma (\mathbf{W_4} \cdot y_t^i + \mathbf{b_4}),
\end{eqnarray}
where $\mathbf{0} = (0, 0, \cdots, 0)$ is a masked exercise embedding with the same dimension as $x_{T+1}$ in Eq.~(\ref{eq:EKTM}). The given input \{$\mathbf{W_3}, \mathbf{W_4}, \mathbf{b_3}, \mathbf{b_4}$\} are the same to those in Eq.~(\ref{eq:EKTM}) without any retraining of EKT.


Moreover, when estimating the knowledge mastery of students by EKT, we can also endow the correspondence between each learnt vector (i.e., in $M$ and $H_t$) and the knowledge concept. Since each vector $H_t^i$ represents the student's state on a certain concept based on the observation of her exercising record at step $t$, we can infer the concept meaning of this vector according to the changes of its value. For example, if we notice that the change of a student's mastery level $l_t^i$ (Eq.~(13), computed by $H_t^i$) is consistent with her exercising score record on concept ``Function'', the corresponding state $H_t^i$ could be viewed as her knowledge state on ``Function'', and correspondingly $M^i$ stores the ``Function'' information. We will conduct the detailed analysis about this estimation in the experiment section.

\begin{figure*}[tp]
	\centering
	\subfigure[Distribution of concepts]{
		\includegraphics[width=0.3\textwidth]{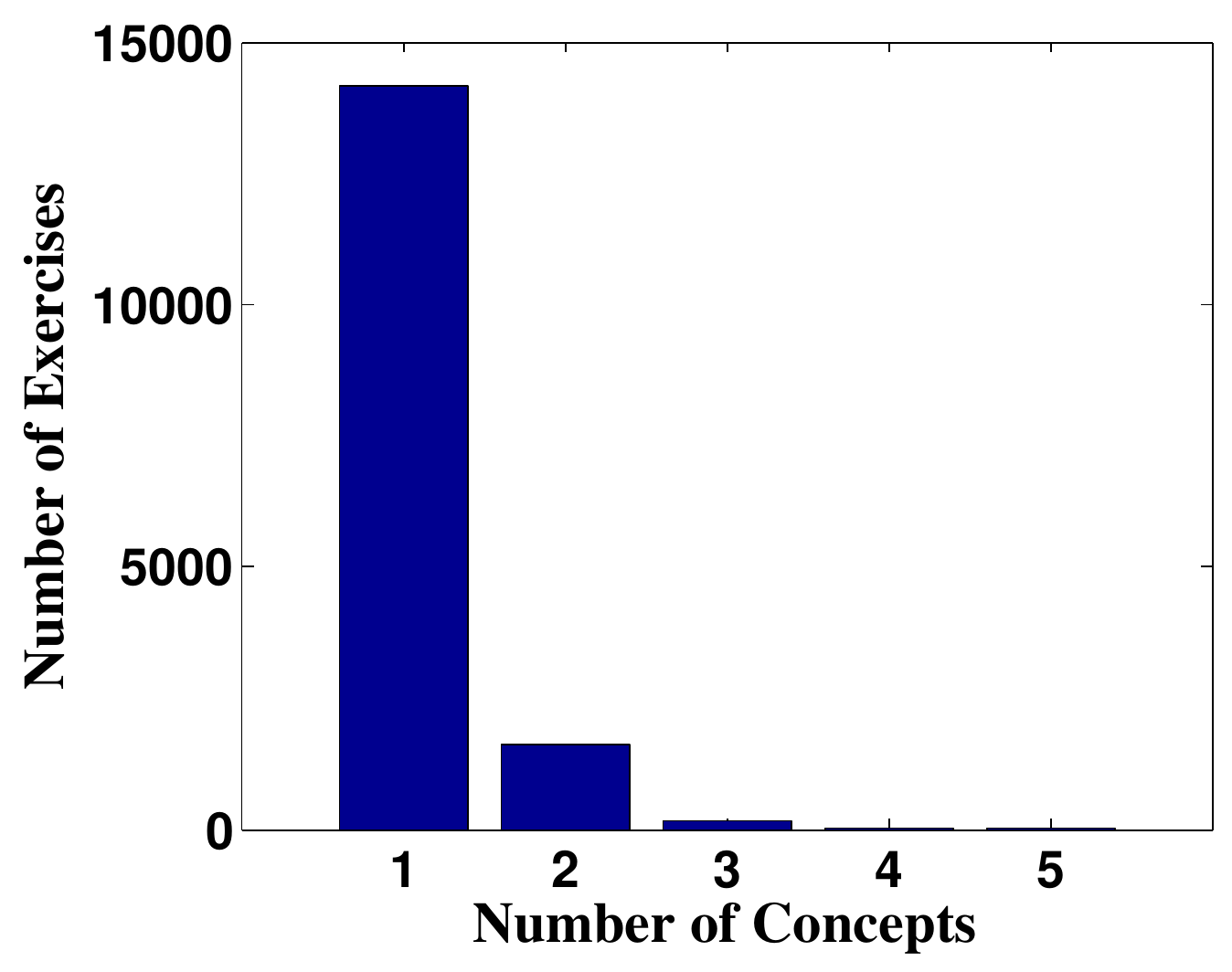}
		\label{fig:subfig:Concept_distribution}
	}
	\subfigure[Distribution of exercising records]{
		\includegraphics[width=0.3\textwidth]{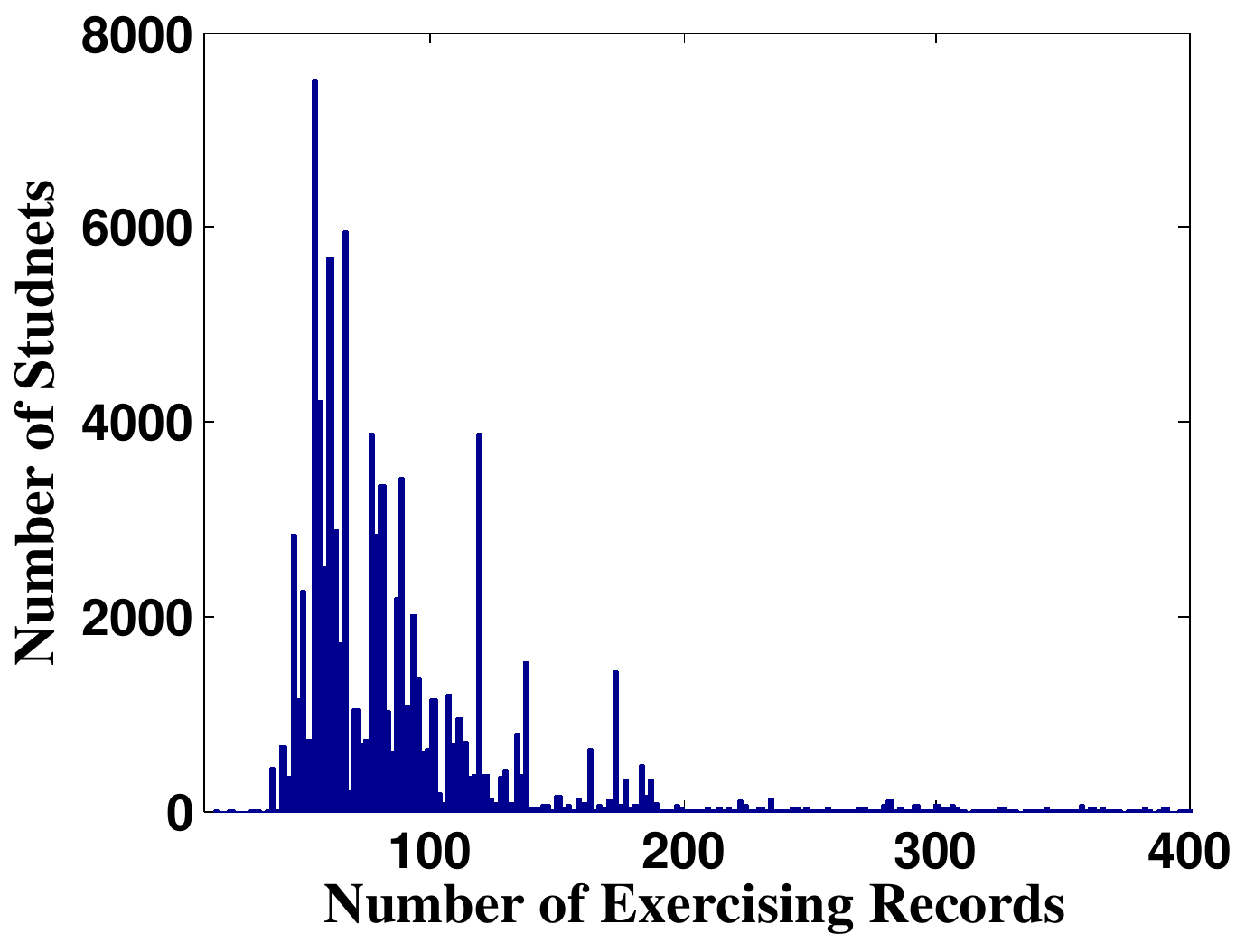}
		\label{fig:subfig:Record_distribution}
	}
	\subfigure[Distribution of content length]{
		\includegraphics[width=0.3\textwidth]{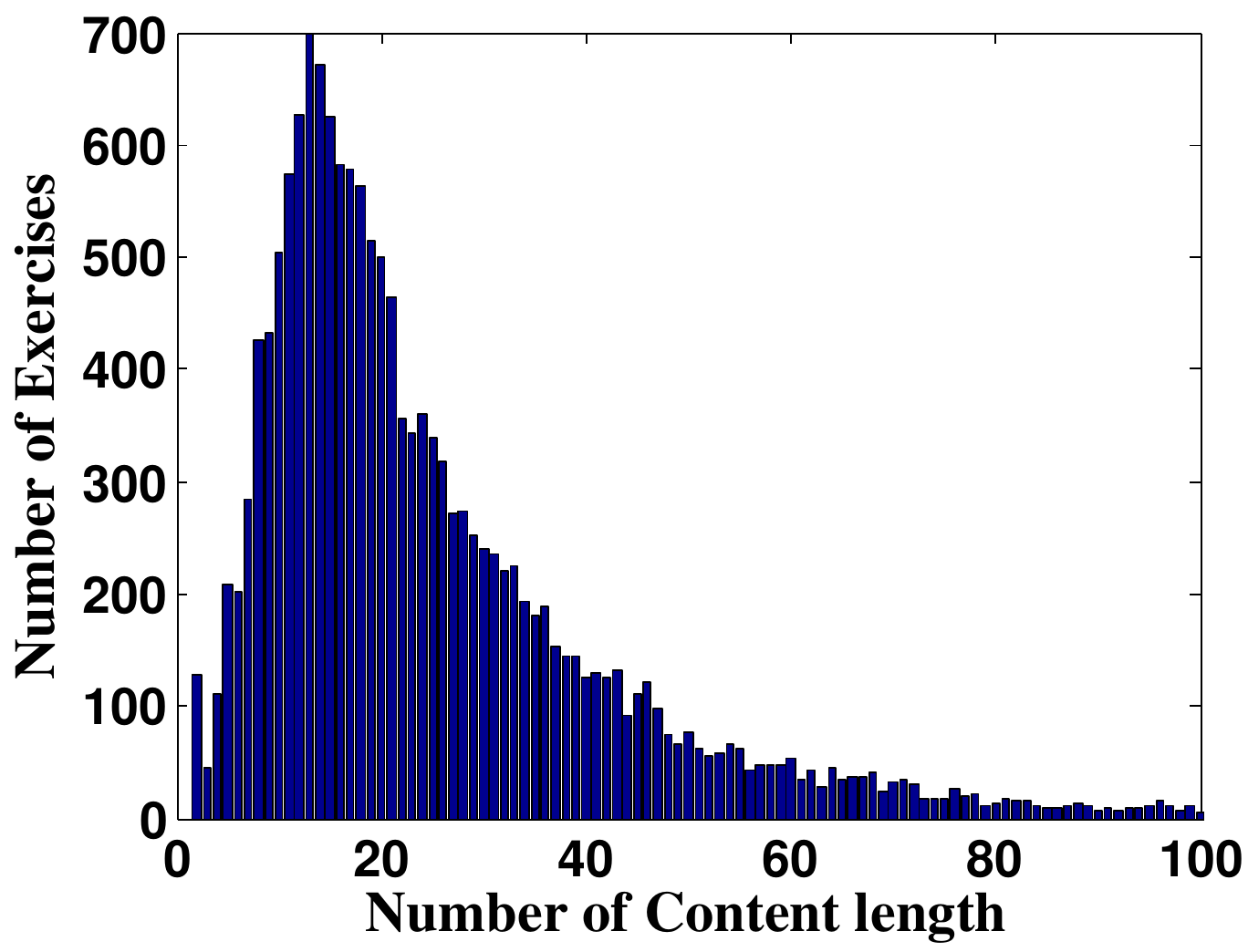}
		\label{fig:subfig:Token_distribution}
	}
	\vspace{-0.3cm}
	\caption{Dataset Analysis: Number distribution of observed data.}
	\label{fig:data_analysis}
	\vspace{-0.6cm}
\end{figure*}

\section{Experiments}
In this section, we conduct extensive experiments to demonstrate the effectiveness of our proposed frameworks and their implementations from various aspects: (1) the prediction performance of EERNN and EKT against the baselines; (2) the effectiveness of attention mechanism in EERNNA and EKTA; (3) the illustration of tracking student's knowledge states by EKT; (4) meaningful visualizations for student performance prediction.

\subsection{Experimental Dataset}
The dataset supplied by iFLYTEK Co., Ltd. was collected from Zhixue\footnote{http://www.zhixue.com}, a widely-used online learning system, which provided high school students with a large number of exercise resources for exercising. In this paper, we conduct experiments on students' records on mathematics because the mathematical dataset is currently the largest and most complete in the system. To make sure the reliability of experimental results, we filtered the students that practiced less than 10 exercises and the exercises that no students had done, and totally, over 5 million exercising records of 84,909 students and 15,045 exercises were remained.

\begin{table}[t]
	\centering
	\caption{The statistics of mathematics dataset.}\vspace{-0.4cm}
	\begin{tabular}{c | c | c}
		\hline
		Statistics & Original & Pruned \\
		\hline
		\# of records & 68,337,149 & 5,596,075 \\
		\# of students & 1,100,726 & 84,909 \\
		\# of exercises& 1,825,767 & 15,045 \\
		\# of knowledge concepts& 37 & 37 \\
		\# of knowledge features& 550 & 447 \\
		Avg. exercising records per student & $\backslash$ & 65.9 \\
		Avg. content length per exercise & $\backslash$ & 27.3 \\
		Avg. knowledge concepts per exercise & $\backslash$ & 1.12 \\
		Avg. knowledge features per exercise & $\backslash$ & 1.8 \\
		Avg. exercises per knowledge concept & $\backslash$ & 406.6\\
		\hline
	\end{tabular}
	\label{tab:dataStats}\vspace{-0.3cm}
\end{table}

\begin{figure}[tp]
	\centering
	\includegraphics[scale=0.9]{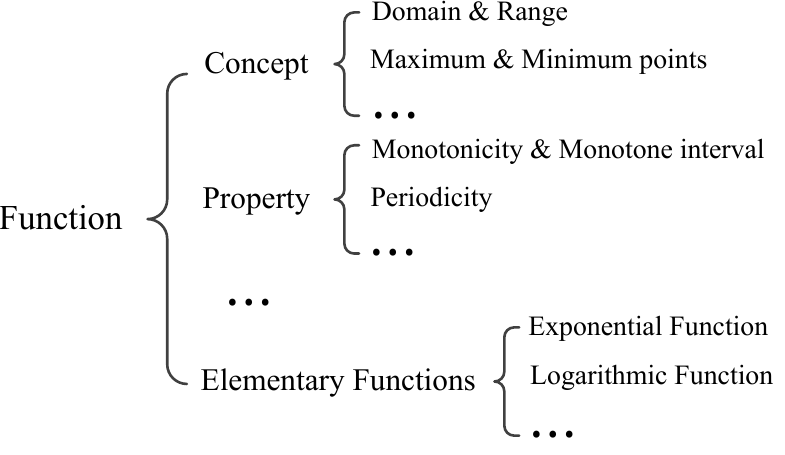}
	\vspace{-0.4cm}
	\caption{An example of the 3-level tree-based structural knowledge system on ``Function'' concept in our dataset. The 1st-level `Function'' totally contains 7 2nd-level concepts and 46 3rd-level concepts. For better illustration, we only show parts of the knowledge system.}
	\vspace{-0.5cm}
	\label{fig:concept_system}
\end{figure}

It is worth mentioning that our dataset contains a 3-level tree-based structural knowledge system labeled by experts, i.e., an explicit hierarchical structure \cite{wang2018exploring}. Thus, each exercise may have multi-level knowledge concepts. Fig.~\ref{fig:concept_system} shows an example of the concept ``Function''. In our dataset, ``Function'' is a 1st-level concept and can be divided into seven 2nd-level sub-concepts (e.g., ``Concept'') and further forty-six 3rd-level sub-concepts (e.g., ``Domain \& Range''). In the following experiments, we treated the 1st-level concepts as the types of knowledge states to be tracked for students in EKT framework and considered all the 2nd-level and 3rd-level sub-concepts as the knowledge features in some baselines (we will discuss later in section 7.2.2). 

We summarized the statistics of the dataset before and after preprocessing in Table~\ref{tab:dataStats}, and also illustrated some data analysis in Fig.~\ref{fig:data_analysis}. Note that most exercises contain less than 2 knowledge concepts and features, and one specific knowledge concept is related to 406 exercises on average. However, the average content length of each exercise is about 27. These observations prove that only using concepts or features cannot distinguish the differences of exercises very well, causing some information loss, and it is necessary to incorporate the exercise content for tracking students' exercising process.


\subsection{Experimental Setup}
In this subsection, we clarify the implementation details to set up our EERNN and EKT frameworks. Then, we introduce the comparison baselines and evaluation metrics in the experiments.

\subsubsection{Implementation Details}
\textbf{Word Embedding.}
The first step is to initialize each word representation for exercise content. Please note that the word embeddings of mathematical exercises in Exercise Embedding are different from traditional ones, like news, because there are some mathematical formulas in the exercise texts. Therefore, to preserve the mathematics semantics, we developed a \emph{formula tool}~\cite{yin2018transcribing} to transform each formula into its \TeX\ code features. For example, the formula ``$\sqrt{x-1}$'' would be the tokens of ``$\backslash$sqrt, \{, x, $-$, 1, \}''. After this initialization, each exercise was transformed into a content sequence with both vocabulary words and \TeX\ tokens. (Fig.~\ref{fig:subfig:Token_distribution} illustrates the distribution of content length of the exercises.) Next, to extract the exclusive word embeddings for mathematics, we constructed a corpus of all 1,825,767 exercises as shown in Table~\ref{tab:dataStats} and trained each word in these exercises into an embedding vector with 50 dimensions (i.e., $d_0$ = 50) by the public \emph{word2vec} tool~\cite{mikolov2013distributed}.

\textbf{Framework Setting.}
We now specify the network initializations in EERNN and EKT. We set the dimension $d_v$ of hidden states in Exercise Embedding as 100, $d_h$ of hidden states in Student Embedding as 100, $d_k$ of knowledge encoding in Knowledge Embedding as 25, and $d_y$ of the vectors for overall presentation in prediction stage as 50, respectively. Moreover, we set the number $K$ of concepts to be tracked in EKT as 37 according to the statistics in Table~\ref{tab:dataStats}.

\textbf{Training Setting.} We followed~\cite{orr2003neural} and randomly initialized all parameters in EERNN and EKT with uniform distribution in the range $(-\sqrt{6/(ni+no)}, \sqrt{6/(ni+no)})$, where $ni$ and $no$ denoted the neuron numbers of feature input and result output, respectively. Besides, we set mini batches as 32 for training and also used dropout (with probability 0.1) to prevent overfitting.

\begin{table}[tp]  
	\centering  
	\fontsize{6.5}{8}\selectfont  
	\caption{Characteristics of all models.}  
	\label{tab:model_characteristic}  	\vspace{-0.4cm}
	\begin{tabular}{c p{0.4cm} p{0.6cm} p{0.6cm} p{0.6cm} c c}  
		\toprule  
		\multirow{2}{*}{Model}& \multicolumn{3}{c}{Data Source}&\multicolumn{2}{c}{Prediction Scenario}& Knowledge\cr  
		\cmidrule(lr){2-4} \cmidrule(lr){5-6} 
		& Score & Concept & Content & General & Cold-start & Tracking? \cr  
		\midrule  
		IRT~\cite{dibello200631a} & \Checkmark & \XSolid & \XSolid & \Checkmark & \XSolid & \XSolid \cr  
		BKT~\cite{corbett1994knowledge} & \Checkmark & \Checkmark & \XSolid & \Checkmark & \XSolid & \Checkmark \cr 
		\midrule  
		PMF~\cite{thai2011factorization} & \Checkmark & \XSolid & \XSolid & \Checkmark & \XSolid & \XSolid \cr  
		DKT~\cite{piech2015deep} & \Checkmark & \Checkmark & \XSolid & \Checkmark & \Checkmark & \XSolid \cr  
		DKVMN~\cite{zhang2017dynamic} & \Checkmark & \Checkmark & \XSolid & \Checkmark &\Checkmark & \Checkmark \cr 
		\midrule
		LSTMM& \Checkmark & \Checkmark & \XSolid & \Checkmark & \Checkmark & \XSolid \cr  
		LSTMA& \Checkmark & \Checkmark & \XSolid & \Checkmark & \Checkmark & \XSolid \cr
		\midrule
		\textbf{EERNNM}~\cite{Su2018exercise}& \Checkmark & \XSolid & \Checkmark & \Checkmark & \Checkmark & \XSolid \cr
		\textbf{EERNNA}~\cite{Su2018exercise}& \Checkmark & \XSolid & \Checkmark & \Checkmark & \Checkmark & \XSolid \cr
		\textbf{EKTM}& \Checkmark & \Checkmark & \Checkmark & \Checkmark &\Checkmark & \Checkmark \cr 
		\textbf{EKTA}& \Checkmark & \Checkmark & \Checkmark & \Checkmark &\Checkmark & \Checkmark \cr 
		\bottomrule  
	\end{tabular}  	\vspace{-0.4cm}
\end{table} 

\begin{figure*} [hbt]
	\centering
	\includegraphics[scale=0.267]{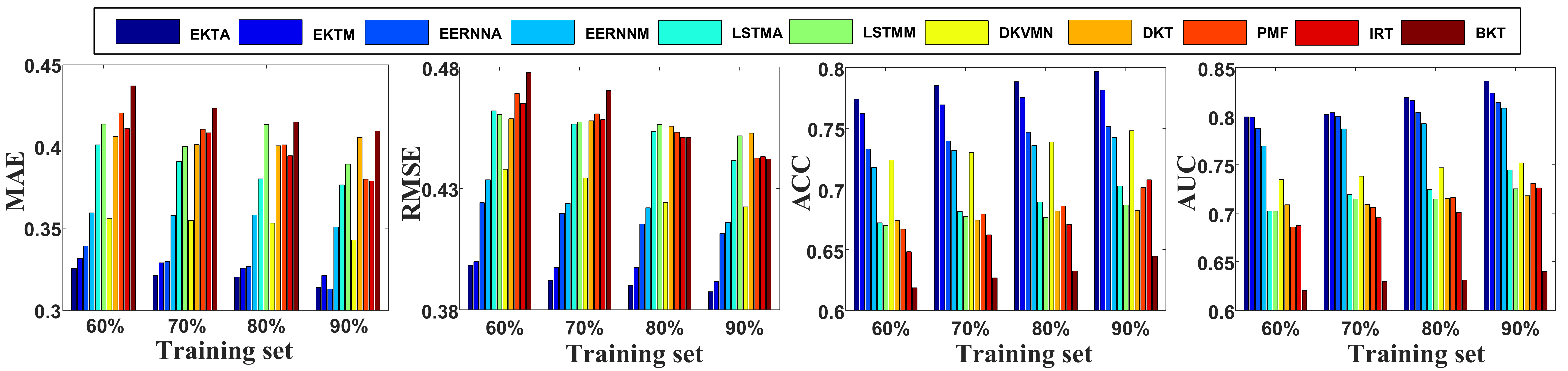}
		\vspace{-0.9cm}
	\caption{Results of student performance prediction in general scenario under four metrics.}
	\label{fig:Overall}
		\vspace{-0.3cm}
\end{figure*}

\begin{figure*} [hbt]
	\centering
	\includegraphics[scale=0.26]{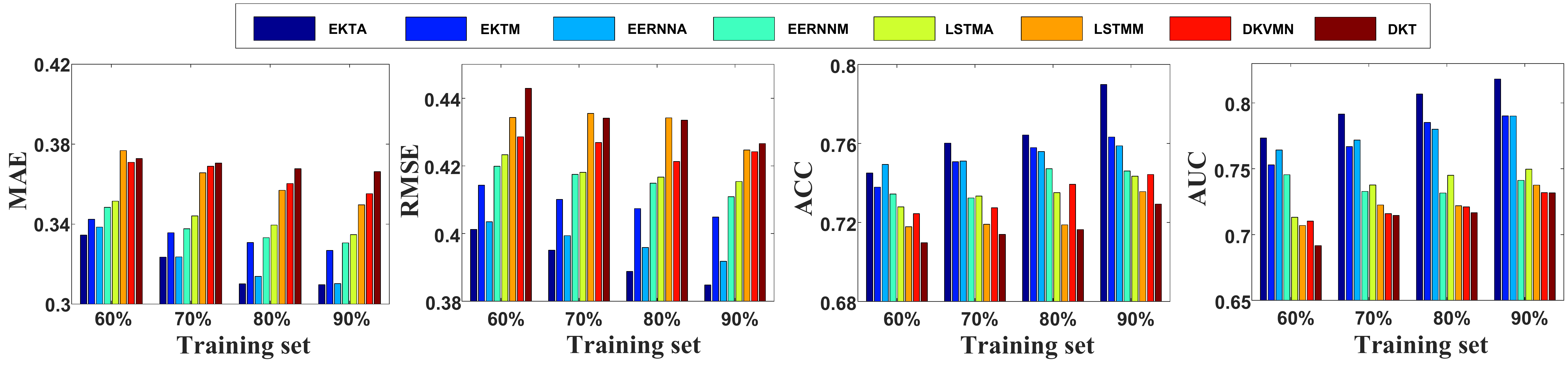}
	\vspace{-0.9cm}
	\caption{Results of student performance prediction on cold-start (new) exercises under four metrics.}
	\vspace{-0.5cm}
	\label{fig:Cold_start}
\end{figure*}

\subsubsection{Comparison Baselines}
To demonstrate the effectiveness of our proposed frameworks, we compared our two EERNN based models, i.e., EERNNM and EERNNA, and two EKT based models, i.e., EKTM and EKTA, with many baselines from various perspectives. Specifically, we chose two models from educational psychology, i.e., \emph{Item Response Theory} (IRT), \emph{Bayesian Knowledge Tracing} (BKT), and three data mining models, i.e., \emph{Probabilistic Matrix Factorization} (PMF), \emph{Deep Knowledge Tracing} (DKT), \emph{Dynamic Key-Value Memory Networks} (DKVMN) for comparison. Then, to highlight the effectiveness of Exercise Embedding in our models, i.e., validating whether or not it is effective to incorporate exercise texts for the prediction, we introduced two variants, which are denoted as LSTMM and LSTMA. The details of them are as follows:
\vspace{-0.15cm}\begin{itemize}
	\setlength{\itemsep}{0pt}
	\setlength{\parsep}{0pt}
	\setlength{\parskip}{0pt}
	\item \textit{IRT}: IRT is a popular cognitive diagnostic model that models student's exercising records by a logistic-like function~\cite{dibello200631a}.
	\item \textit{BKT}: BKT is a typical knowledge tracing model which assumes the knowledge states of each student as a set of binary variables and traces them separately with a kind of hidden Markov model~\cite{corbett1994knowledge}.
	\item \textit{PMF}: PMF is a factorization model that projects students and exercises into latent factors~\cite{thai2011factorization}.
	\item \textit{DKT}: DKT is a deep learning method that leverages recurrent neural network (RNN and LSTM) to model students' exercising process for prediction~\cite{piech2015deep}. The inputs are the one-hot encodings of student-knowledge representations.
	\item \textit{DKVMN}: DKVMN is a state-of-the-art deep learning method that could track student states on multiple concepts~\cite{zhang2017dynamic}. It contains a \emph{key} matrix to store concept representation and a \emph{value} matrix for each student to update the states. However, it does not consider the effect of exercise content in the modeling.
	\item \textit{LSTMM}: LSTMM is a variant of EERNN framework. Here, in the modeling process, we do not embed exercises from their contents, and only represent them as the one-hot encodings with both 2nd-level and 3rd-level knowledge features\footnote{The one-hot representation is a typical manner in many models. We use knowledge features for representation because the number of them is much larger than the 1st-level ones, ensuring the reliability.}. Then we leverage traditional LSTM to model students' exercising process. For prediction, LSTMM follows Markov property strategy similar to EERNNM.
	\item \textit{LSTMA}: LSTMA is another variant of EERNN framework which contains the same modeling process as LSTMM. For prediction, LSTMA follows the strategy of Attention mechanism similar to EERNNA.
\end{itemize}\vspace{-0.15cm}

For better illustration, we list the detailed characteristics of these models in Table~\ref{tab:model_characteristic}. More specifically, in the experiments, we used the open source to implement the BKT model\footnote{https://github.com/IEDMS/standard-bkt}, and all other models were implemented by ourselves by PyTorch~\cite{paszkepytorch} using Python on a Linux server with four 2.0GHz Intel Xeon E5-2620 CPUs and a Tesla K20m GPU. All models were tuned to have the best performance to ensure the fairness.

\subsubsection{Evaluation Metrics}
A qualified model for student performance prediction should have good results from both regression and classification perspectives. In this paper, we evaluated the prediction performance of all models using four widely-used metrics in the domain~\cite{fogarty2005case,wu2015cognitive,wu2017knowledge,zhang2019hierarchical,kuang2018stable}.

From the regression perspective, we selected \emph{Mean Absolute Error} (MAE) and \emph{Root Mean Square Error} (RMSE), to quantify the distance between predicted scores and the actual ones. The smaller the values are, the better the results have. Besides, we treated the prediction problem as a classification task, where an exercising record with score 1 (0) indicates a positive (negative) instance. Thus, we used two metrics, i.e., \emph{Prediction Accuracy} (ACC), \emph{Area Under an ROC Curve} (AUC), for measuring. Generally, the value 0.5 of AUC or ACC represents the performance prediction result by randomly guessing, and the larger, the better.

\subsection{Student Performance Prediction}
\textbf{Prediction in General Scenario.} In this subsection, we compare the overall performance of all models on student performance prediction. To set up the experiments, we partitioned the dataset from student's perspective, where the exercising records of each student are divided into training set and testing set with different percentages. Specifically, for a certain student, we used her first 60\%, 70\%, 80\%, 90\% exercising records (with the exercises she practiced and the scores she got) as training sets, and the remains were for testing, respectively. We repeated all experiments 5 times and report the average results using all metrics.

Fig.~\ref{fig:Overall} shows the overall results on this task. There are several observations. First, all our proposed EKT based models and EERNN based models perform better than other baseline methods. The results clearly indicate that both EKT and EERNN frameworks can make full use of both exercising records and exercise materials, benefiting the prediction performance. Second, among our proposed models, we find that EKT based models (EKTA, EKTM) generate better results than EERNN based ones (EERNNA, EERNNM), indicating the effectiveness of tracking student's knowledge states on multiple concepts ($H_t$ in Fig.~(\ref{fig:EKT_model_architecture})) than simply modeling them with an integrated encoding ($h_t$ in Fig.~(\ref{fig:EERNN_model_architecture})). Third, models with Attention mechanism (EKTA, EERNNA, LSTMA) outperform those with Markov property (EKTM, EERNNM, LSTMM), which demonstrates that it is effective to track the focused student embeddings based on similar exercises for the prediction. Next, as our proposed models incorporate an independent Exercise Embedding module for extracting exercise encoding directly from the text content, they outperform their variants (LSTMA, LSTMM) and the state-of-the-arts (DKVMN, DKT). This observation also suggests that both EKT and EERNN alleviate the information loss caused by the feature-based or knowledge-specific representations in existing methods. Last but not least, the traditional models (IRT, PMF and BKT) do not perform as well as deep learning models in most cases. We guess a possible reason is that these RNN based deep models can effectively capture the change of student's exercising process, and therefore, the deep neural network structures are suitable for student performance prediction.

In summary, all above evidences demonstrate that both EKT and EERNN have a good ability to predict student performance by taking full advantage of both the exercising records and exercise materials. Moreover, EKT shows the superiority of tracking student's multiple knowledge states for the prediction.

\begin{figure}[tp]
	\centering
	\subfigure[ACC Comparison]{
		\includegraphics[width=0.23\textwidth]{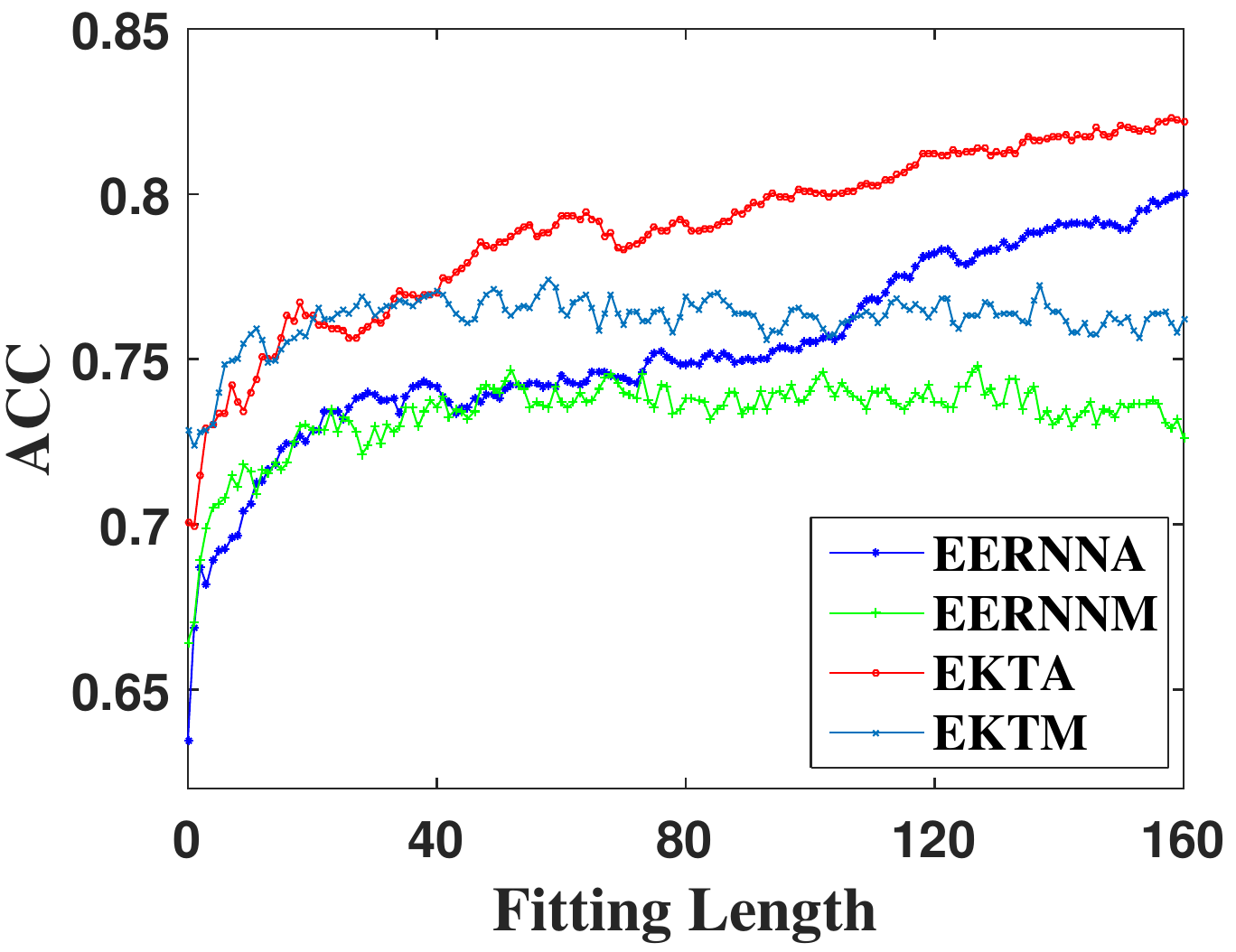}
		\label{fig:subfig:Seq_ACC}
	}
	\subfigure[AUC Comparison]{
		\includegraphics[width=0.23\textwidth]{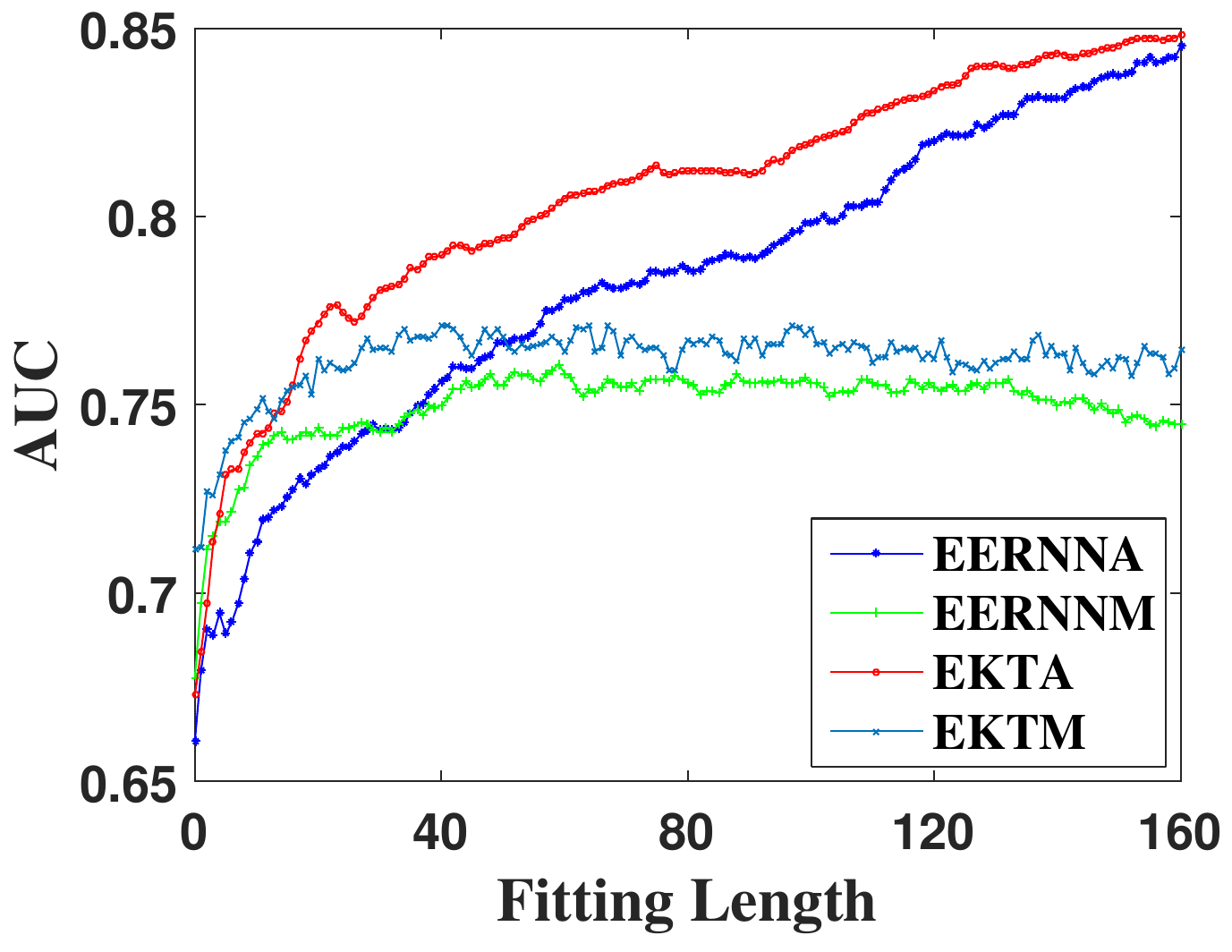}
		\label{fig:subfig:Seq_AUC}
	}
	\vspace{-0.3cm}
	\caption{The effectiveness of attention in fitting process for testing.}
	\label{fig:Seq_influence}
	\vspace{-0.6cm}
\end{figure}

\textbf{Prediction on Cold-start (new) Exercises.} The task of predicting student performance often suffers from the ``cold start" problem. Thus, in this part, we conduct detailed experiments to demonstrate the performance of our proposed models in this scenario from the exercise's perspective (Experimental analysis on the cold-start students will be given in the following subsection). Specifically, we selected the new exercises (that never show up in training) in our experiment. Then we only trained each model on 60\%, 70\%, 80\%, 90\% training sets, and tested the prediction results on these new exercises in the corresponding testing sets. Please note that, in this experiment, we did not change any training process and just selected the cold-start exercises for testing, thus all the models do not need any retraining. 

For better illustration, we reported the experimental results of all deep learning based models under all metrics in Fig.~\ref{fig:Cold_start}. There are also similar observations as Fig.~\ref{fig:Overall}, which demonstrate the effectiveness of both EKT and EERNN frameworks once again. Clearly, from the results, EKT based models, especially EKTA, perform the best, followed by EERNN based models. Also, we find that the improvement of them for prediction on new exercises are more significant. Thus, we can reach a conclusion that both EKT and EERNN with Exercise Embedding module for representing exercises from the text content could effectively distinguish the characteristics of each exercise. Those models are superior to LSTM based models of using feature representation as well as the state-of-the-art DKVMN and DKT of considering knowledge representation. In summary, both EKT and EERNN can deal with the cold-start problem when predicting student performance on new exercises. 

\begin{figure}[tp]
	\centering
	\subfigure[EERNNA]{
		\includegraphics[width=0.23\textwidth]{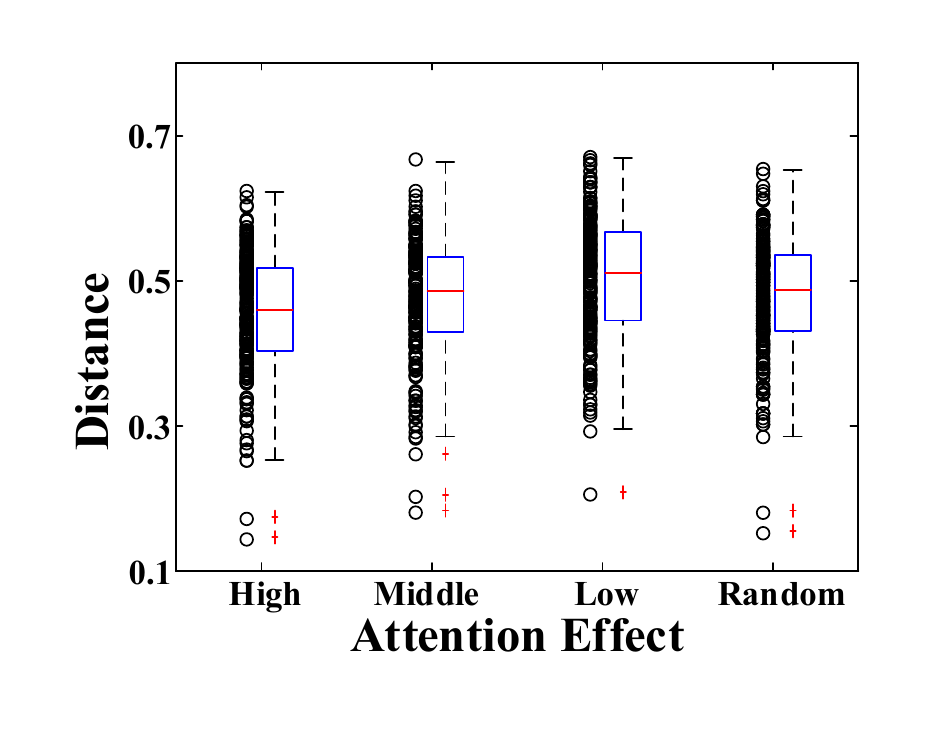}
		\label{fig:subfig:EERNN_attn}
	}
	\subfigure[EKTA]{
		\includegraphics[width=0.23\textwidth]{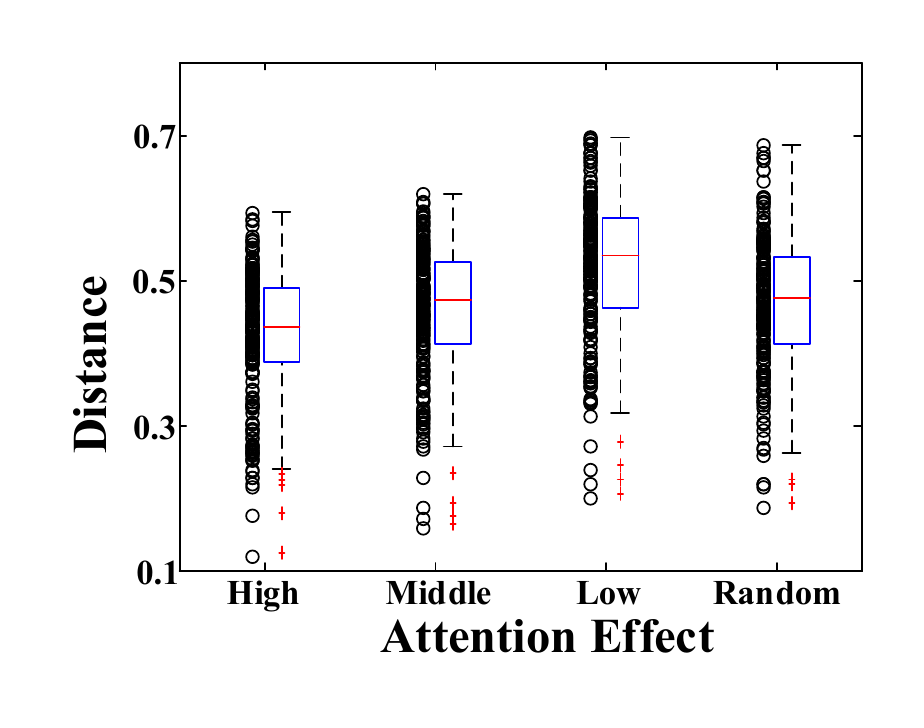}
		\label{fig:subfig:EKT_attn}
	}
	\vspace{-0.3cm}
	\caption{Performance over different attention values in proposed models.}
	\label{fig:Attn_effect}
	\vspace{-0.6cm}
\end{figure}	

\begin{figure*} [hbt]
	\centering
	\includegraphics[scale=0.28]{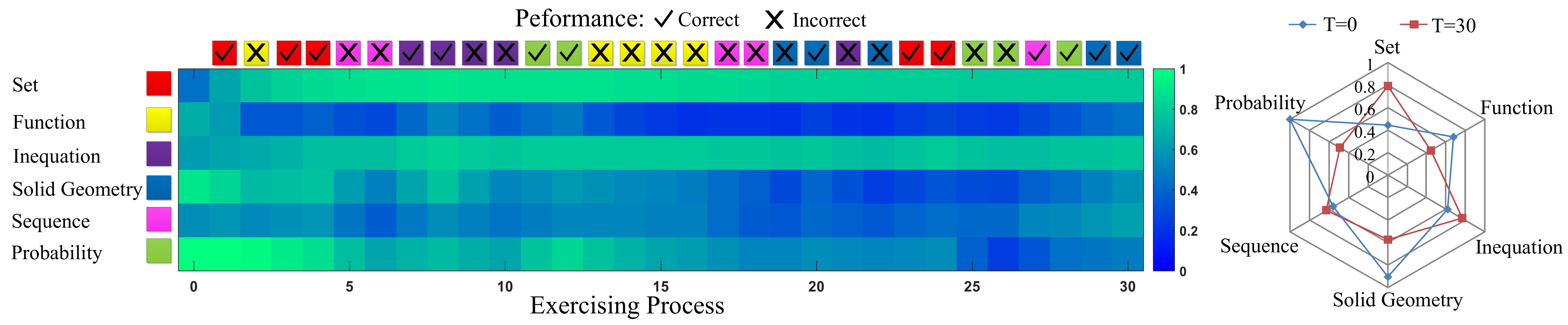}
	\vspace{-0.5cm}
	\caption{An example of the knowledge mastery level tracking of a certain student on 6 concepts during her 30 exercising steps, which is painted in the middle matrix. Left side shows all concepts, which are marked in different colors. Top line records her performance on the 30 exercises. Right radar figure shows her knowledge mastery levels (in the range $(0, 1)$) on 6 concepts before (T=0) and after (T=30) exercising.}
	\label{fig:Tracking}
	\vspace{-0.3cm}
\end{figure*}

\subsection{Effectiveness of Attention}

As we have clarified in EKT and EERNN, \emph{EKTA (EERNNA) with Attention mechanism} has a superior ability than EKTM (EERNNM) because the former ones can track the focused student states and enhance the effect of these states when modeling each student's exercising process. To highlight the effectiveness of attention, we compared the performance of our proposed models, i.e., EKTA (EERNNA) and EKTM (EERNNM). To setup this experiment, we first divided the students into 90\%/10\% partitions, using the 90\% students for training and the remaining 10\% for testing. Therefore, the testing students never showed up in training. Then, for each student in the testing process, we fitted her exercising sequence by the trained models with different length step $t$ from 0 to 180 and predicted her scores on the last 10\% exercises of her exercising records. We also conducted 10-fold cross validation to ensure the reliability of experimental results. Here, we reported the average performance under ACC and AUC metrics.

Fig.~\ref{fig:subfig:Seq_ACC} and Fig.~\ref{fig:subfig:Seq_AUC} show the comparison results of them. From the figures, all models perform better and better as the length of fitting sequence increases. Specifically, for EERNNA and EERNNM, we find that they generate similar results when the fitting sequence of student is short (less than 40), however, as the fitting length increases, EERNNA performs better gradually. When the length surpasses about 60, EERNNA outperforms EERNNM significantly. Moreover, we also clearly see that both EKTA and EKTM outperform EERNNA and EERNNM on both metrics, respectively. Based on this phenomenon, we can draw the following conclusions. First, both EKTM and EERNNM are effective at the beginning of student's exercising but discards some important information when the sequence is long. Comparatively, EKTA and EERNNA enhance the effect of some of student's historical states with the attention mechanism, benefiting the prediction. Second, EKT framework has better prediction ability by incorporating the information of knowledge concepts into modeling, which is superior to EERNN. Third, notice that our proposed EKT (EERNN) based models obtain about 0.72 (0.65) on the metrics of both ACC and AUC (much better than the randomly guessing 0.5), by the prior student state $H_0$ in EKT (Fig.~\ref{fig:EKT_model_architecture}) and $h_0$ in EERNN (Fig.~\ref{fig:EERNN_model_architecture}), in the case of predicting the first performance of new students without any record (i.e., the fitting length is 0). Moreover, they all get better predictions with more fitting records even if the sequence is not very long at the first few steps. This finding also demonstrates that both EKT and EERNN based models can guarantee the performance in the cold-start scenario when making prediction for new students.

One step further, we also show the effectiveness of \emph{EKTA (EERNNA) with Attention mechanism} with detailed analysis from a data correlation perspective, i.e., we could get better prediction results based on the higher attention score (i.e., $\alpha$ in Eq.~(12) and Eq.~(6)). Specifically, for predicting the performance of a certain student at one specific testing step (e.g., the score on $e_{T+1}$), we first computed and normalized the attention scores of her historical exercises (i.e., $\{e_1, e_2, \cdots, e_T\}$) calculated by EKTA (EERNNA) into [0, 1]. Then, we partitioned these exercises into the low ([0, 0.33]), middle ((0.33, 0.66]) and high ((0.66, 1]) groups based on attention scores. In each group (e.g., the low), the average response score of the student on these exercises were used to represent the response score of this group. Then, for all testing steps of the specific student, we computed and illustrated the Euclidean Distance between the response scores in each group (i.e., the low, middle, high) and the scores for prediction (i.e., the scores on $\{e_{T+1}, e_{T+2}, \cdots\}$). Finally, Fig.~\ref{fig:Attn_effect} illustrates the distance results of all students in the forms of both scatter and box figures. At each time step, we also added a result computed with a group of 10 randomly selected exercises (namely, Random) for better illustration. From the figure, in both EKT and EERNN models, the response scores of the exercises in high attention groups have smallest distances (large correlation) with the score for prediction while the low groups are farthest. This finding demonstrates that the higher the attention value is, the more contribution of this exercise will make when predicting the response score on a new exercise. In conclusion, both EKT and EERNN frameworks can improve the prediction performance by incorporating the attention mechanism.

\begin{figure*}[t]
	\centering
	\includegraphics[scale=0.5]{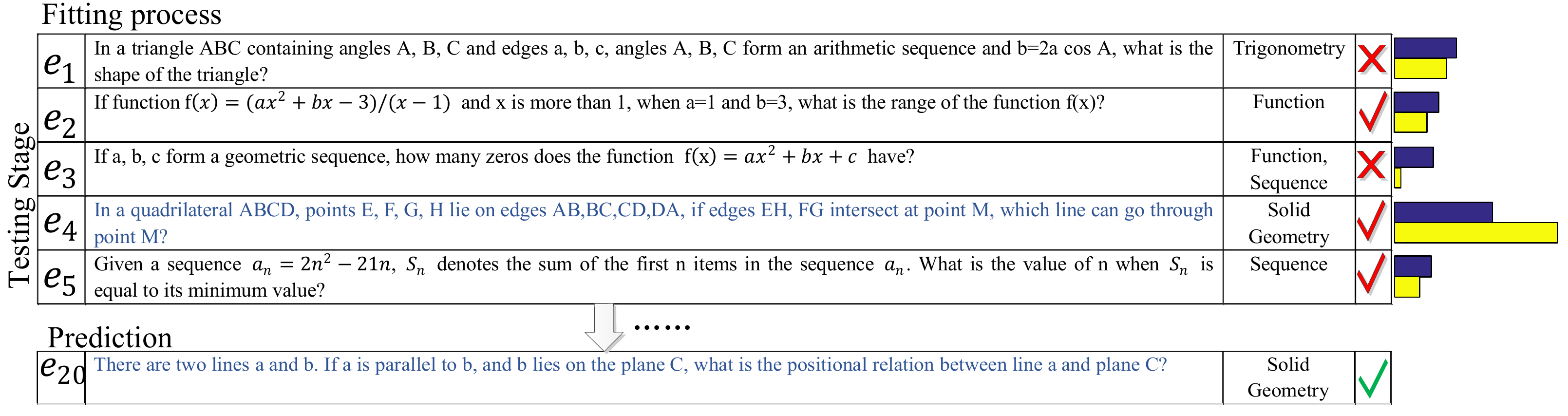}
	\vspace{-0.5cm}
	\caption{\small{Attention visualization in EERNNA and EKTA of an example student. We predict her performance on $e_{20}$ based on her past 19 exercise records (we only show the first 5 exercises for better illustration). Right bars show the attention scores of two frameworks (i.e., EERNNA (blue) and EKTA (yellow)) for all exercises based on $e_{20}$.}}
	\label{fig:case}
	\vspace{-0.45cm}
\end{figure*}

\subsection{Visualizations}
\textbf{Visualization of Knowledge Acquisition Tracking.} The important ability of EKT, which is superior to EERNN, is that it can track student's knowledge states on multiple concepts to further explain the change of knowledge mastery levels of the student, ensuring the interpretability. To make deep analysis about this claim, we visualize the predicted mastery levels (i.e., $l_t^i$ in Eq.~(\ref{eq:mastery})) of a certain student on explicit knowledge concepts at each step during the exercising process. For better visualization, we made some preprocessing as follows. First, we selected 6 most frequent concepts that the student practiced since it was hard to illustrate clearly if we visualize all 37 concepts in one figure. Second, we just logged students' performance records on the knowledge concepts rather than distinguishing each specific exercise. In other words, if the student correctly answered an exercise about ``Function'', we logged that she answered ``Function'' right. Then, we visualized the change of her states on these concepts modeled by EKTA (as a representative).

Fig.~\ref{fig:Tracking} shows the throughout results. In the left of this figure, the first column means the initial mastery levels of this student (i.e., $H_0$ at $T$=$0$ in Fig.~\ref{fig:EKT_model_architecture}) on 6 concepts without any exercising, where her states differs from each other. Then, she starts exercising with the following 30 exercises on these concepts. Meanwhile, her states  on the concepts (output by EKTA) change gradually during the steps. Specifically, when she answers an exercise right (wrong), her knowledge state on the corresponding concept increases (decreases), e.g., she acquires knowledge on ``Set'' after she solves an exercise of ``Set'' concept at her second exercising step. During her exercising process, we can see that she gradually masters the concept ``Set'' but is incapable of understanding ``Function'' since she does all exercises on ``Set'' right but fails to solve all exercises on ``Function''. However, there exists an inconsistent phenomenon that her mastery level of ``Function'' becomes slightly lower at the third exercising step even she answers the exercise correctly. This is because the model may not perfectly track the student with only few exercising records at the beginning, but it could get better performance if the student's exercising records are getting longer enough in the following steps. As a result of her 30 exercises, we can explain that at last this student has well mastered the concepts of ``Set'' and ``Inequation'', partially mastered ``Solid Geometry'', ``Sequence'' and ``Probability'', but failed on ``Function'', as illustrated in the right radar figure.

\textbf{Visualization of Student Performance Prediction.} Both EERNNA and EKTA also have great powers of explaining the prediction results by the attention mechanism (i.e., the attention score $\alpha$ in Eq.~(\ref{equ:hatt}) and Eq.~(\ref{equ:Hatt})). As an example, Fig.~\ref{fig:case} illustrates the attention scores for a student's exercises. Here, both EERNNA and EKTA predict that the student can answer exercise $e_{20}$ correctly, because she got right answers on a similar exercise $e_4$ in the past. Let us take into consideration about the exercise materials, we can conclude: (1) $e_4$ is actually much more difficult than $e_{20}$; (2) both $e_{20}$ and $e_4$ contain the same knowledge concept ``Solid Geometry''. In addition, we notice that EKTA endows a larger attention weight on $e_4$ than EERNNA, since EKTA can incorporate the exercise concepts into the modeling. This visualization clearly hints that both EKTA and EERNNA are able to provide good ways for analyzing and explaining the prediction results, which is quite meaningful in real-world applications.

\section{Conclusions}
 In this paper, we presented a focused study on student performance prediction. Specifically, we first proposed a general \emph{E}xercise-\emph{E}nhanced \emph{R}ecurrent \emph{N}eural \emph{N}etwork (EERNN) framework exploring both student's exercising records and the content of corresponding exercises. Though EERNN could effectively deal with the problem of predicting student performance on future exercises, it can not track student's knowledge states on multiple explicit concepts. Therefore, we then extended EERNN to an \emph{E}xercise-aware \emph{K}nowledge \emph{T}racing (EKT) framework by further incorporating the information of knowledge concepts existed in each exercise. For making final predictions, we designed two strategies under both EKT and EERNN, i.e., straightforward \emph{EKTM (EERNNM) with Markov property} and sophisticated \emph{EKTA (EERNNA) with Attention mechanism}. Comparatively, EKTA (EERNNA) could track the historically focused
 information of students for making prediction, which was superior to EKTM (EERNNM). Finally, we conducted extensive experiments on a large-scale real-world dataset, and the results demonstrated the effectiveness and interpretability of our proposed models.

\section*{Acknowledgements}
	This research was partially supported by grants from the National Natural Science Foundation of China (Grant No.s 61672483, U1605251， and 91746301), the Science Foundation of Ministry of Education of China \& China Mobile (No. MCM20170507), and the Iflytek joint research program. Qi Liu gratefully acknowledges the support of the Young Elite Scientist Sponsorship Program of CAST and the Youth Innovation Promotion Association of CAS (No. 2014299). Zhenya Huang would like to thank the China Scholarship Council for their support.

\begin{tiny}
	\bibliographystyle{abbrv}
	\bibliography{TKDE}
\end{tiny}

\begin{IEEEbiography}[{\includegraphics[width=1in,height=1.25in,clip,keepaspectratio]{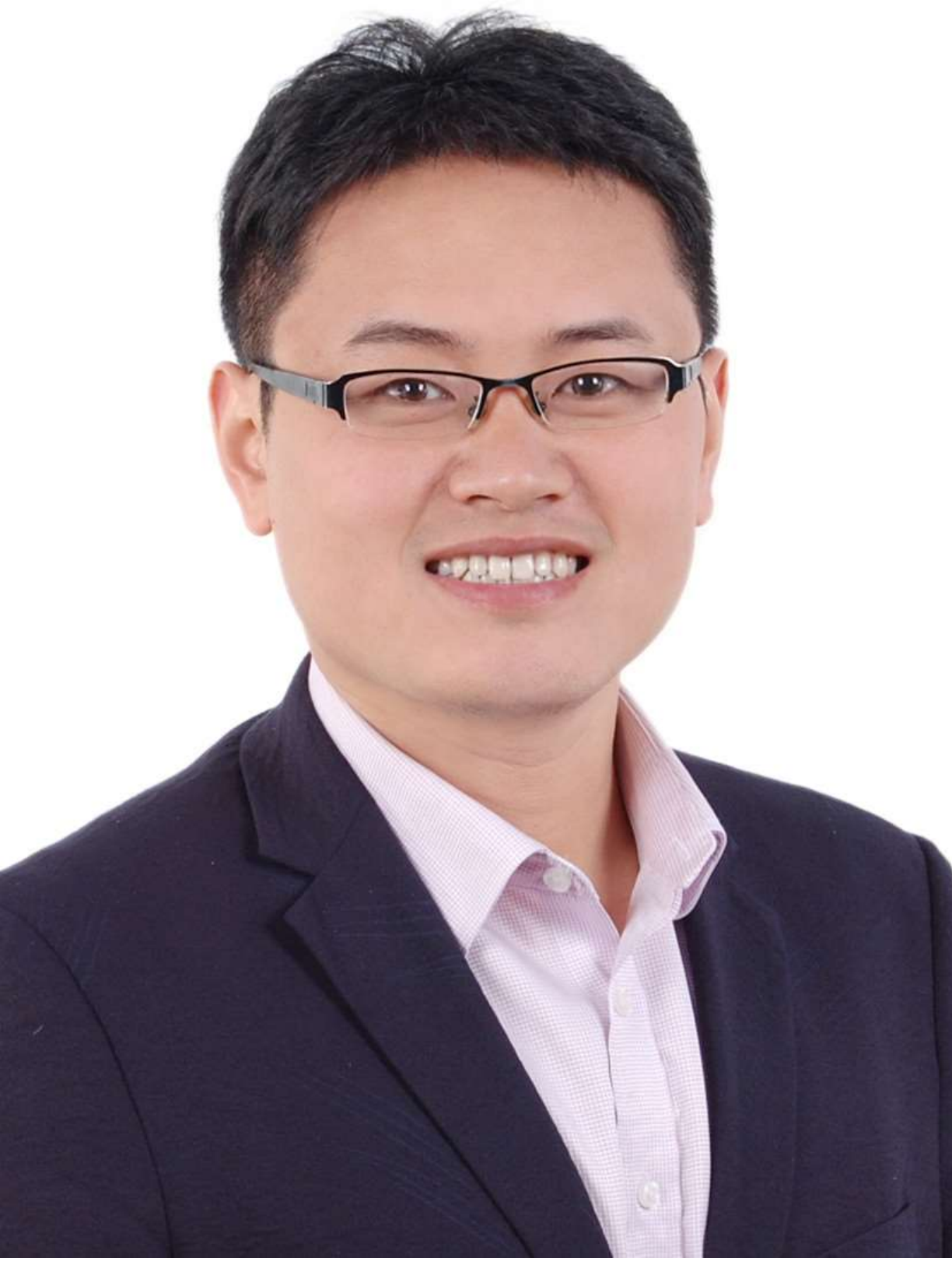}}]
	{Qi Liu} is an associate professor at University of Science and Technology of China (USTC). He received the  Ph.D. degree in Computer Science from USTC. His general area of research is data mining and knowledge discovery. He has published prolifically in refereed journals and conference proceedings, e.g., TKDE, TOIS, TKDD, TIST, KDD, IJCAI, AAAI, ICDM, SDM and CIKM. He has served regularly in the program committees of a number of conferences, and is a reviewer for the leading academic journals in his fields. He is a member of ACM and IEEE. Dr. Liu is the recipient of the KDD 2018 Best Student Paper Award (Research) and the ICDM 2011 Best Research Paper Award. He is supported by the Young Elite Scientist Sponsorship Program of CAST and the Youth Innovation Promotion Association of CAS. 
\end{IEEEbiography}

\begin{IEEEbiography}[{\includegraphics[width=1in,height=1.25in,clip,keepaspectratio]{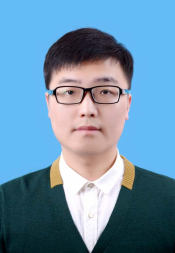}}]
	{Zhenya Huang} received the BE degree in software engineering from Shandong University (SDU), China, in 2014. He is currently working toward the Ph.D. degree in the School of Computer Science and Technology at University of Science and Technology of China (USTC). His main research interests include data mining and knowledge discovery, recommender systems and intelligent education systems. He has published several papers in referred conference proceedings, such as AAAI'2016, AAAI'2017, CIKM'2017, KDD'2018, AAAI'2018 and DASFAA'2018.
\end{IEEEbiography}

\begin{IEEEbiography}[{\includegraphics[width=1in,height=1.25in,clip,keepaspectratio]{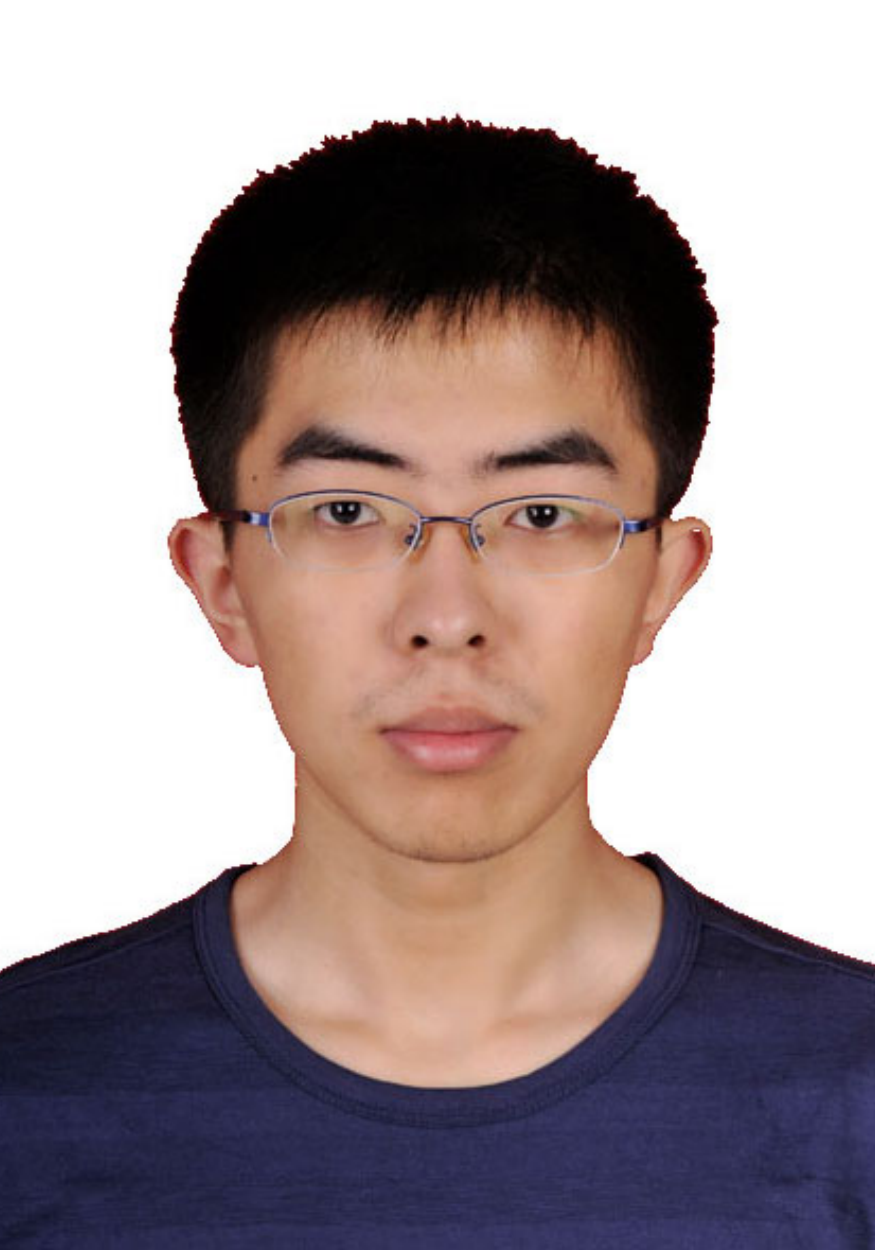}}]
	{Yu Yin} received the BE degree in computer science from University of Science and Technology of China (USTC), China, in 2017. He is currently working toward the ME degree in the School of Computer Science and Technology at University of Science and Technology of China (USTC). His main research interests include data mining, intelligent education systems, image recognition. He won the first prize in the Second Student RDMA Programming Competition, 2014. He has published papers in referred conference proceedings, such as AAAI'2018 and KDD'2018.
\end{IEEEbiography}

\begin{IEEEbiography}[{\includegraphics[width=1in,height=1.25in,clip,keepaspectratio]{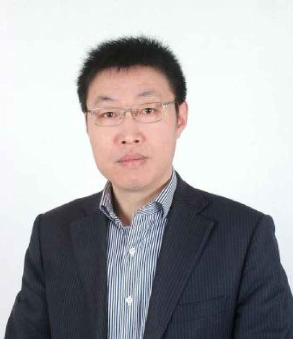}}]
	{Enhong Chen}(SM'07) is a professor and vice dean of the School of Computer Science at University of Science and Technology of China  (USTC). He received the Ph.D. degree from USTC. His general area of research includes data mining and machine learning, social network analysis and recommender systems. He has published more than 100 papers in refereed conferences and journals, including IEEE Trans. KDE, IEEE Trans. MC, KDD, ICDM, NIPS, and CIKM. He was on program committees of numerous conferences including KDD, ICDM, SDM. He received the Best Application Paper Award on KDD-2008, the Best Student Paper Award on KDD-2018 (Research), the Best Research Paper Award on ICDM-2011 and Best of SDM-2015. His research is supported by the National Science Foundation for Distinguished Young Scholars of China.
	He is a senior member of the IEEE.
\end{IEEEbiography}

\begin{IEEEbiography}[{\includegraphics[width=1in,height=1.25in,clip,keepaspectratio]{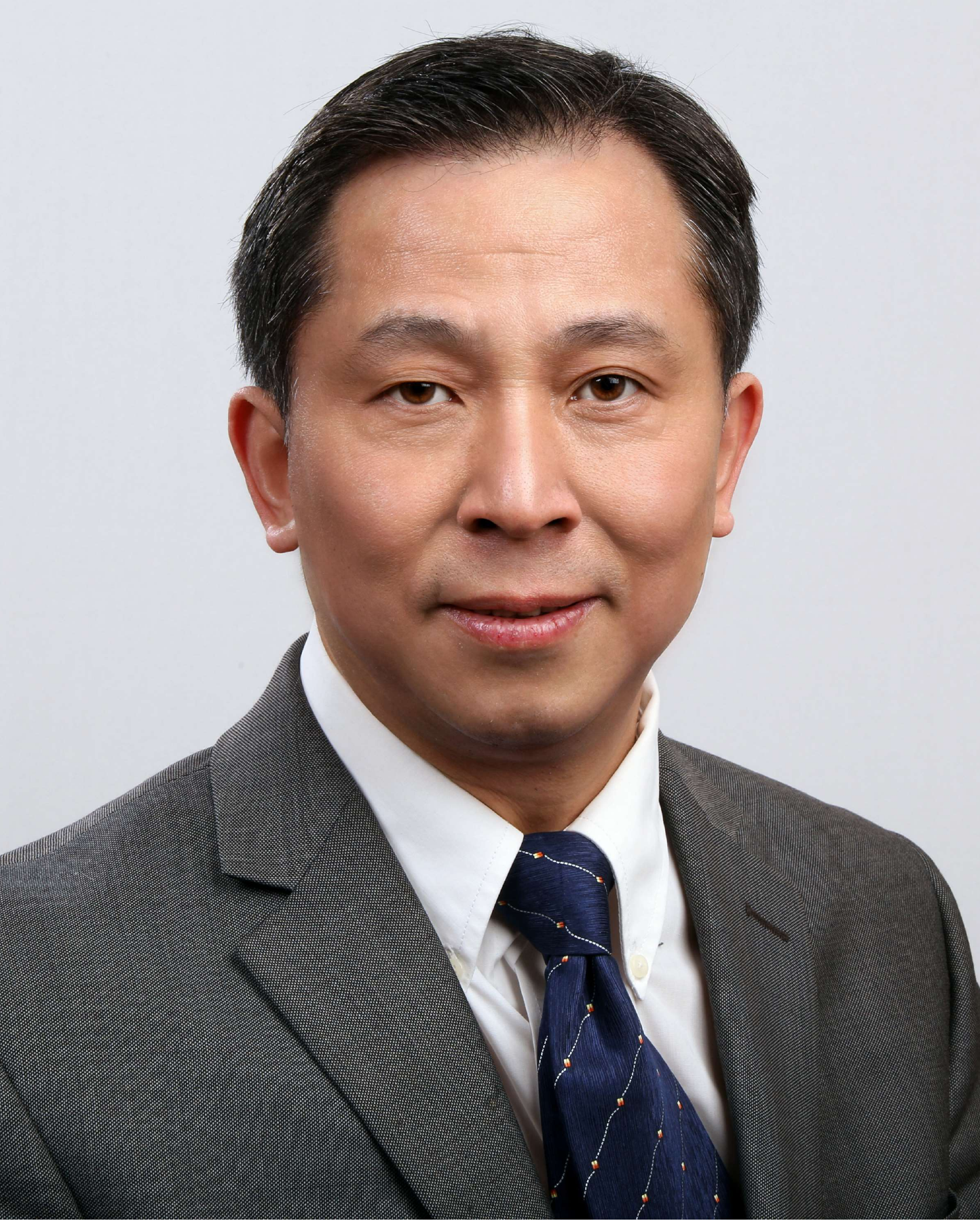}}]
	{Hui Xiong} (SM'07) is currently a Full Professor at Rutgers, the State University of New Jersey, where he received the ICDM-2011 Best Research Paper Award, and the 2017 IEEE ICDM Outstanding Service Award. His general area of research is data and knowledge engineering, with a focus on developing effective and efficient data analysis techniques for emerging data intensive applications. He has published prolifically in refereed journals and conference proceedings (4 books, 80+ journal papers, and 100+ conference papers). He is a co-Editor-in-Chief of Encyclopedia of GIS, an Associate Editor of IEEE TKDE, IEEE TBD, ACM TKDD, and ACM TMIS. He has served regularly on the organization and program committees of numerous conferences, including as a Program Co-Chair of the Industrial and Government Track for KDD-2012, a Program Co-Chair for ICDM-2013, a General Co-Chair for ICDM-2015, and a Program Co-Chair of the Research Track for KDD-2018. For his outstanding contributions to data mining and mobile computing, he was elected an ACM Distinguished Scientist in 2014.
\end{IEEEbiography}

\begin{IEEEbiography}[{\includegraphics[width=1in,height=1.25in,clip,keepaspectratio]{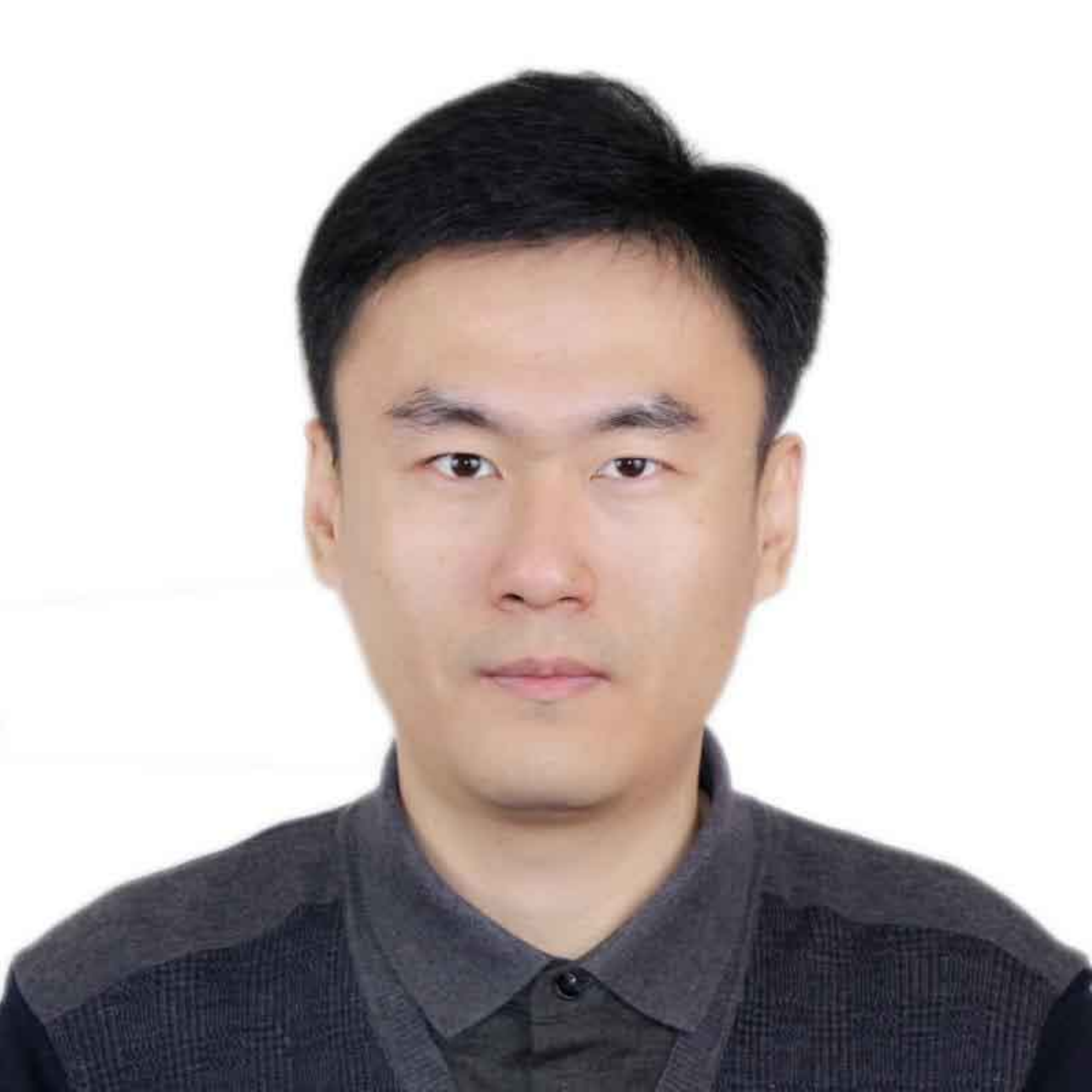}}]
	{Yu Su} is a researcher of IFLYTEK CO., LTD. He received the Ph.D. degree from Anhui University. His main area of research includes data mining, machine learning, recommender systems and intelligent education systems. He has published several papers in referred conference proceedings and journals, such as IJCAI'2015, AAAI'2017, AAAI'2018, KDD'2018, CIKM'2017, DASFAA'2016, ACM TIST.
\end{IEEEbiography}

\begin{IEEEbiography}[{\includegraphics[width=1in,height=1.25in,clip,keepaspectratio]{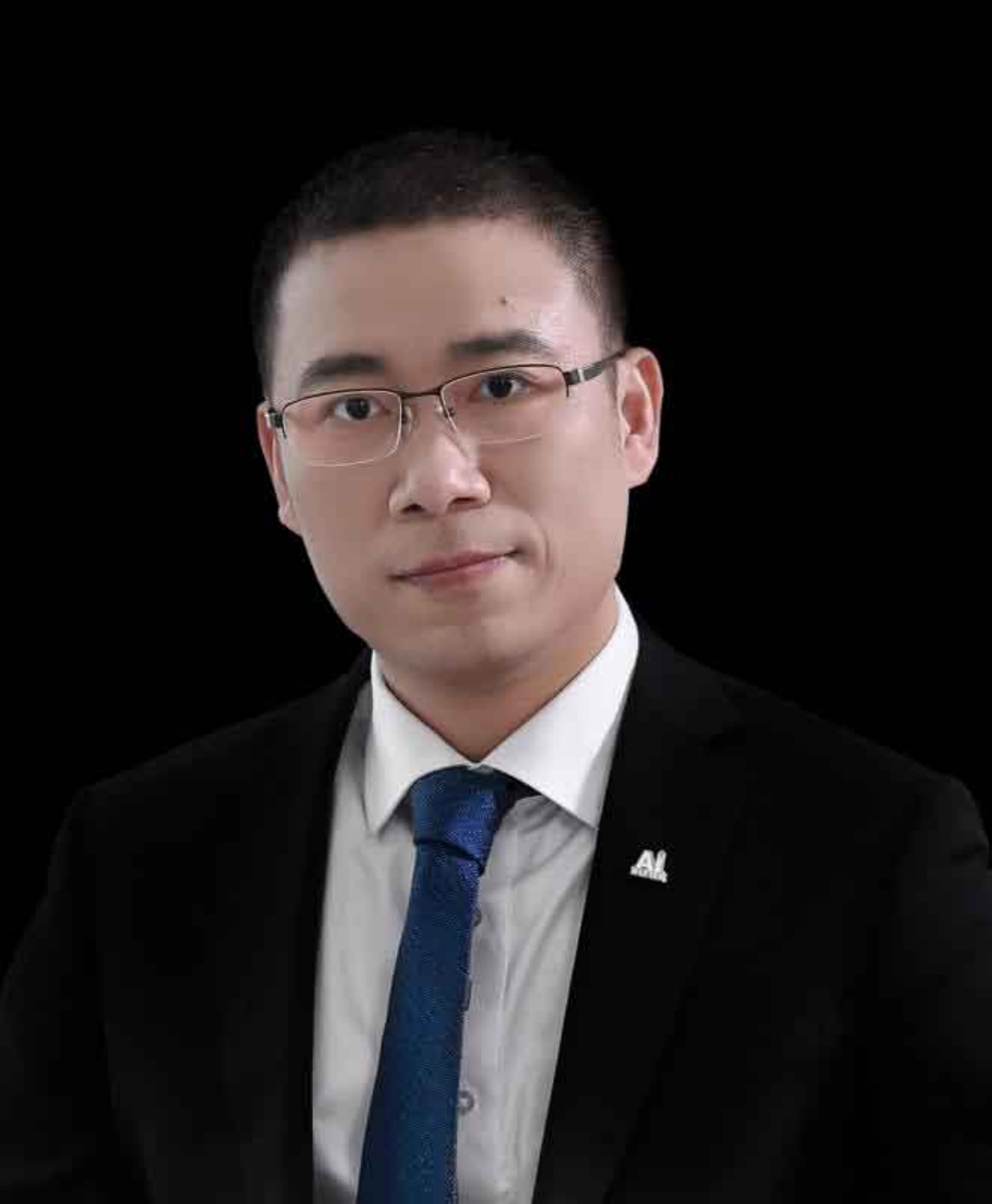}}]
	{Guoping Hu} received the Ph.D. degree from University of Science and Technology of China. He is one of the founders of iFLYTEK Co. Ltd. Currently, he is senior vice president of iFLYTEK, dean of Research Institute of iFLYTEK, director of National Key Laboratory of Cognitive Intelligence, vice chairman of the Strategic Alliance of New Generation AI Industrial Technology Innovation, and deputy group leader of the Overall Group of National AI Standardization. He has possessed 65 invention patents, and published over 20 papers in core journals and important international conferences at home and abroad.
\end{IEEEbiography}

\end{document}